\documentclass[lineno]{jfm}
\usepackage{graphicx}
\usepackage{epsfig}
\usepackage{amsmath}
\usepackage{booktabs}
\usepackage{natbib}
\usepackage{hyperref}
\usepackage{soul}
\usepackage{accents}
\hypersetup{
	colorlinks = true,
	urlcolor   = blue,
	citecolor  = black,
}

\usepackage{graphicx}
\usepackage[arrowdel]{physics}
\usepackage{gensymb}
\usepackage{amsfonts}
\usepackage{caption}
\usepackage{subcaption}
\usepackage[dvipsnames]{xcolor}
\usepackage[normalem]{ulem}

\newcommand\Rh{{\text{Rh}}}

\newcommand{\avg}[1]{\langle{#1}\rangle}

\newcommand{\ks}[1]{\textcolor{black}{#1}}
\newcommand{\vd}[1]{\textcolor{black}{#1}}
\newcommand{\vbj}[1]{\textcolor{black}{#1}}
\newcommand{\red}[1]{\textcolor{black}{#1}}
\newcommand{\new}[1]{\textcolor{black}{#1}}

\newcommand{\nnew}[1]{\textcolor{black}{#1}}
\newcommand{\revi}[1]{\textcolor{black}{#1}}
\newcommand{\vdd}[1]{\textcolor{black}{#1}}

\newcommand{\final}[1]{\textcolor{black}{#1}}

\title{Cascades transition in generalised two-dimensional turbulence}

\author{Vibhuti Bhushan Jha\aff{1,2}\corresp{\email{vibhutibjha@gmail.com}},
	Kannabiran Seshasayanan\aff{2}\corresp{\email{kanna@iitm.ac.in}}
	\and \\ 
	Vassilios Dallas\aff{3,4}\corresp{\email{vassilios.dallas@gmail.com}}}

\affiliation{\aff{1}Space Applications Centre, Indian Space Research Organisation, Ahmedabad, Gujarat, India
	\aff{2}Department of Applied Mechanics, Indian Institute of Technology Madras, Chennai, India
	\aff{3}Mathematical Institute, University of Oxford, Oxford, OX2 6GG, UK
	\aff{4}Environmental Research Laboratory, National Centre for Scientific Research ``Demokritos'', 15341 Athens, Greece
}

\begin{document}
	\maketitle
	
	\begin{abstract}
		\final{Generalised} two-dimensional (2D) fluid \revi{dynamics} is characterised by a relationship between a scalar field $q$, called generalised vorticity, and the stream function $\psi$, 
		namely \revi{$q = (-\nabla^2)^\frac{\alpha}{2} \psi$}.
		We study the transition of cascades in generalised 2D turbulence by systematically varying the parameter $\alpha$ and investigating its influential role in determining the directionality (inverse, forward, or bidirectional) of these cascades. 
		We derive upper bounds for the dimensionless dissipation rates of generalised energy $E_G$ and enstrophy $\Omega_G$ as the Reynolds number tends to infinity. These findings corroborate numerical simulations, illustrating the inverse cascade of $E_G$ and forward cascade of $\Omega_G$ for $\alpha > 0$, contrasting with the reverse behaviour for $\alpha < 0$. The dependence of dissipation rates on system parameters reinforces these observed transitions, substantiated by spectral fluxes and energy spectra, which hint at Kolmogorov-like scalings at large scales but discrepancies at smaller scales \revi{between numerical and theoretical estimates}. These discrepancies are possibly due to nonlocal transfers, which dominate the dynamics as we go from positive to negative values of $\alpha$. 
	Intriguingly, \revi{the} forward cascade of $E_G$ for $\alpha < 0$ reveals similarities to three-dimensional turbulence, notably the emergence of vortex filaments within a 2D framework, marking a unique feature of this generalised model.
	\end{abstract}
	
	\begin{keywords}
		Keywords
	\end{keywords}

\section{Introduction}
The presence of nonlinear interactions \vbj{allows} energy to be 
\vdd{transferred} across different length scales in turbulent flows. While in three-dimensional (3D) turbulence, energy is 
\vdd{transferred} to small scales (forward cascade), energy is transferred to large scales in two-dimensional (2D) turbulence (inverse cascade). \revi{This, along with the existence of a second quadratic invariant, makes 2D turbulence an exciting system.} Not limited to theoretical or numerical \vbj{studies}, such behaviour has been observed in atmospheric \citep{lilly1969numerical,rhines1979geostrophic,nastrom1984kinetic,read2001transition} and planetary flows \citep{siegelman2022moist}. 
\vd{The presence of external mechanisms like rotation, stratification, magnetic field or compactification of one dimension can make the prediction of the cascade direction challenging \citep{celani2010turbulence,alexakis2011two,sen2012anisotropy,marino2013inverse,deusebio2014dimensional}. 
This is due to the competing behaviour between the forward and inverse cascades of the quadratic invariants, often dependent on some control parameter like anisotropy, rotation rate, etc.} \vdd{This results in bidirectional cascades of the quadratic invariants.}
Such bidirectional cascades have been observed in numerical \citep{Smithetal1996,alexakis2011two,Seshasayananetal2014,sozza2015dimensional} 
and experimental settings \citep{shats2010turbulence,xia2011upscale,campagne2014direct}. Furthermore, there is evidence for \vd{bidirectional cascades} in atmospheric \citep{byrne2013height,king2015upscale,shao2023physical}, \revi{oceanic \citep{ScottWang2005,Arbicetal2013,Balwadaetal2016,Khatrietal2018, balwada2022direct}} and planetary flows \citep{lesur2011non,young2017forward}. 
	
The transition from one type of cascade to another can be either smooth \revi{or occur at a critical point \citep{AlexakisBiferale2022, alexakis2023quasi, van2024phase}}. 
The presence of a critical dimension at which the cascade direction changes has been demonstrated by \cite{frisch1976crossover} \revi{ and recently examined further by \cite{verma2024critical}}. 
\cite{benavides2017critical} found the transition from a forward to a bidirectional cascade \vd{of energy} in thin layers of fluid turbulence to be critical \citep{Ecke2017}.
In 2D magnetohydrodynamic (MHD) turbulence, the variation in magnetic field \vbj{strength} above a critical value leads to cascades transition \vd{of the quadratic invariants} \citep{Seshasayananetal2014}, and the \vbj{inverse} energy flux scales as a power law. Similarly under certain regimes, MHD flows can behave like a 2D flow at large scales and 3D flow at smaller scales \citep{alexakis2011two}. In this case, the inverse cascade of energy is highly sensitive to the strength of the magnetic field.

\cite{Pierrehumbertetal1994} considered a generalised model of 2D turbulence characterised by a parameter $\alpha$ which links the streamfunction to a scalar field called generalised vorticity. 
Certain values of $\alpha$
lead to equations relevant in the context of geophysical flows. 
For $\alpha = 2$\revi{,} the generalised model gives the familiar barotropic vorticity equation from 2D Navier-Stokes \citep{Tabeling2002,BoffettaEcke2012}. For $\alpha = 1$\revi{,} it gives the surface quasi-geostrophic (SQG) equation, which describes the motion of a rotating stratified fluid \citep{Heldetal1995,Lapeyre2017}. For $\alpha=-2$\revi{,}  it gives 
a rescaled shallow-water quasi-geostrophic equation in the asymptotic limit of length scales large compared to the deformation scale \citep{LarichevMcWilliams1991,Smithetal2002}. 
In generalised 2D turbulence, changing the value of $\alpha$ leads to varying degrees of forward and inverse cascades \vd{of the quadratic invariants}. 
There are various studies using different values of $\alpha$ and analysing the associated energy spectra \citep{tran2004nonlinear,watanabe2004unified,Iwayamaetal2015}.
		
	\vd{Two dimensional instabilities can couple the forced modes to large scale 2D modes, transferring energy to larger scales, which indicate the possibility of an inverse cascade \citep{alexakis2018three}. This nonlocal interaction of scales can be quantified with the notion of negative eddy viscosity \citep{Kraichnan1976}}.  
	The cascades of 2D Navier-Stokes turbulence ($\alpha = 2$) have been predicted to be less local than for 3D turbulence \citep{Kraichnan1971, BoffettaEcke2012}. 
	\cite{Pierrehumbertetal1994} focus on positive values of $\alpha$ only and argue that $\alpha=2$ is at the transition between local to nonlocal transfers \vd{of generalised enstrophy}, in the sense that dominant straining of small scales comes from the largest scales when $\alpha > 2$.
\vd{On the other hand, \cite{WatanabeIwayama2007,Foussardetal2017} found nonlocal interactions in the generalised enstrophy transfer for $1 \leq \alpha \leq 2$.
For the energy transfers \cite{Pierrehumbertetal1994} argued that \vbj{it becomes } nonlocal when $\alpha > 4$.}

The value of $\alpha$ decides the nature of the cascades. For positive $\alpha$, energy (enstrophy) cascades to large (small) scales, whereas energy (enstrophy) cascades to small (large) scales for negative $\alpha$. Thus, depending on the sign and value of $\alpha$, the cascade would either be inverse, forward or bidirectional. \revi{This suggests that there is a transition in the direction of the cascades as one changes the value of $\alpha$. It is also to be noted that the aforementioned studies have not analysed} systematically the $\alpha < 0$ regime. 

In this work we focus on the numerical study of the cascades in generalised 2D turbulence by systematically varying $\alpha$ in the range $-1 \leq \alpha \leq 2$. While it may be possible to extend the runs to values of $\alpha < -1$, it is computationally challenging to achieve a statistically stationary regime in the range of parameters we have considered in this study.
We perform \vdd{direct numerical simulations (DNS)} to determine the transition of the cascades and we use exact mathematical inequalities to derive bounds on generalised energy and enstrophy dissipation rates. Finally, our study deals with the degree of non-locality of triadic interactions \vdd{by analysing shell-to-shell transfers in the range $-1 \leq \alpha \leq 2$.}
	
The paper is organised as follows. In section \ref{sec:problem} \vdd{we discuss the problem formulation.} 
\vd{We present the numerical results in section \ref{sec:results} and in section \ref{sec:bounds_discussion} we discuss the theoretical bounds on the generalised energy and enstrophy dissipation rates that we derive in appendix \ref{sec:bounds}. Finally, we summarise our concluding remarks in section \ref{sec:conclusion}.}

\section{Problem Formulation}\label{sec:problem}
	Consider the 2D evolution equation of the generalised vorticity $q(x,y,t)$ in a periodic domain \vdd{of size} $[0,2\pi]\times[0,2\pi]$,
 \vbj{	\begin{equation}
		\partial_t q + \revi{J(q, \psi)} = -\nu_+(-\nabla^2)^n q -\nu_-(-\nabla^2)^{-m} q + f_q,
		\label{eq:qmodel}
	\end{equation}}
where $\psi(x,y,t)$ is the streamfunction, the nonlinear term \revi{is given by the Jacobian $J(q, \psi) = \partial_x q \partial_y \psi - \partial_x \psi \partial_y q$,} 
$\nu_+$ is the \vdd{hyper-viscous coefficient, $\nu_-$ is the hypo-viscous coefficient } and $f_q$ is the external forcing.
The fluid velocity is given by $\bold u = \nabla \times (\psi \hat z)$. The  generalised relationship between $q$ and $\psi$ in 2D fluid dynamics is,
\revi{
	\begin{equation}
		q = (-\nabla^{2})^{\frac{\alpha}{2}} \psi
		\label{eq:psiq}
	\end{equation}
    }
	as discussed by \revi{\cite{Pierrehumbertetal1994,Smithetal2002,tran2004nonlinear,TeitelbaumMininni2012}}.
	
In the presence of an inverse cascade the energy of the large-scale modes grow to extreme values forming a condensate, the growth saturates when the viscous dissipation at the largest scale balances the energy injection \citep{Tabeling2002,Chanetal2012}. To prevent the formation of a very large condensate and to reach a turbulent stationary regime we supplement our system with a large scale dissipative term \vd{$-\nu_- (-\nabla^2)^{-m} q$} that is responsible for saturating the inverse cascade. 
\vd{To model very high Reynolds number flows we consider a hyper-viscous term $-\nu_+ (-\nabla^2)^{n} q$ by raising the Laplacian to an integer power $n > 1$. This term provides a wider inertial range, as it \final{sets} in at much smaller scales compared to the normal viscous term (with $n = 1$).}
The \final{hypo-viscous term similarly extends the} inertial range for scales larger than the forcing scale. 
	
	We denote the dimensions of a quantity by using the square bracket $[\cdot]$. Considering the fact that the variables $x, y$ have  units  of  length  $[L]$, and $t$ has  units  of  time $[T]$, then $[\psi] = [L]^2[T]^{-1}$ and $[q] = [L]^{2-\alpha} [T]^{-1}$. \revi{Using \eqref{eq:psiq} we define $f_\psi$, the forcing acting on the stream function equation with $f_q$ as, $f_q = (-\nabla^{2})^{\frac{\alpha}{2}} f_\psi$. Taking $f_0$ to be the amplitude of $f_\psi$,} the two dimensionless control parameters are the small scale Reynolds number 
\begin{equation}
	Re_+ = \sqrt{\vbj{f_0}k_f} / (\nu_+k_f^{2n-1/2}), 
\end{equation} 
and the large scale Reynolds number 
\begin{equation}
	Re_- = \sqrt{\vbj{f_0}k_f} / (\nu_-k_f^{-2m-1/2}), 
\end{equation}
with $[f_0] = [L]^2[T]^{-2}$, $[\nu_+] = [L]^{2n}[T]^{-1}$ and $[\nu_-] = [L]^{-2m}[T]^{-1}$.
	
In the limit of $\nu_+ \to 0$, $\nu_- \to 0$ and $f_q = 0$ the integral over a periodic domain of any function of the scalar field $q$ is \nnew {formally} conserved. Therefore, there are infinite number of invariants \citep{Smithetal2002}. Following the Kolmogorov-Kraichnan phenomenology of 2D Navier-Stokes turbulence, the two quadratic invariants that determine the cascade directions in generalised 2D turbulence are:
\revi{
\begin{equation}
 E_G = \frac{1}{2}\avg{\psi q}, \quad \Omega_G = \frac{1}{2}\avg{q^2},
\end{equation}
}
where we refer to $E_G$ as the generalised energy and $\Omega_G$ as the generalised enstrophy with units $[E_G] = [L]^{4-\alpha}[T]^{-2}$ and $[\Omega_G] = [L]^{4-2\alpha}  [T]^{-2}$. The \revi{angular} brackets $\avg{\cdot}$ denote spatiotemporal averaging. Multiplying \eqref{eq:qmodel} by $\psi$ and integrating over space and time, we can derive the evolution equation of the generalised energy. \revi{Setting \vbj{$dE_G/dt = 0$} due to statistical stationarity we get}
\revi{
\begin{equation}
\epsilon  =  \epsilon_+ + \epsilon_-, \label{eq:energybal}
\end{equation}}
where  $\epsilon_+ = \nu_+\avg{\psi (-\nabla^{2})^{n}q}$ is the small scale dissipation rate, $\epsilon_- = \nu_-\avg{\psi (-\nabla^{2})^{-m}q}$ is the large scale dissipation rate and $\epsilon = \avg{\psi f_q}$ is the injection rate of the generalised energy. Similarly, if we multiply \eqref{eq:qmodel} with $q$ and integrate over space and time, we obtain the evolution equation of the generalised enstrophy. \revi{Setting \vdd{$d\Omega_G/dt = 0$} due to statistical stationarity we get}, 
\revi{
	\begin{equation}
		\xi  = \xi_+ + \xi_-,  \label{eq:enstrophybal}
	\end{equation}}
	where $\xi_+ = \nu_+\avg{q (-\nabla^{2})^nq}$ is the small scale dissipation rate, $\xi_- =  \nu_-\avg{q (-\nabla^{2})^{-m}q}$ is the large scale dissipation rate and $\xi = \avg{q f_q}$ is the injection rate of the generalised enstrophy.
	
	The equivalent expression of \eqref{eq:psiq} in Fourier space is 
    \revi{
	\begin{equation}
		\widehat{q} ({\bf k}, t) = k^\alpha \widehat{\psi} ({\bf k}, t)
		\label{eq:coupling}
	\end{equation} 
    }
	where $k = \sqrt{k_x^2 + k_y^2}$ is the isotropic two-dimensional wavenumber. The notation $\widehat{\cdot}$ denotes the Fourier transform coefficients. \revi{The spectra} of the two quadratic invariants can be connected using the relationship \eqref{eq:coupling} to get
	\begin{equation}
		\Omega_G(k) = k^\alpha E_G(k).
		\label{eq:connection}
	\end{equation}
	Now, the spectra of the two quadratic invariants in the inertial range can be derived following \cite{K41} scaling arguments, where the spectral flux is assumed to be constant in the inertial range and the spectral densities $E_G(k)$ and $\Omega_G(k)$ are only functions of the local scale and spectral flux. Hence, using dimensional arguments for the \nnew{generalised energy}, we get
 \revi{
	\begin{align}
		& \frac{k E_G(k)}{\tau_E(k)} = \epsilon = const., \quad 
		\tau_E(k) = [k^{5 - \alpha} E_G(k)]^{-1/2},
		\label{eq:fluxEG} \\
		& \frac{k \Omega_G(k)}{\tau_\Omega(k)} = \xi = const., \quad 
		\tau_\Omega(k) = [k^{5 - 2\alpha}\Omega_G(k)]^{-1/2},
		\label{eq:fluxOG}
	\end{align}
}
where $\tau_E(k)$ is the local \vbj{turnover} timescale that it takes to transfer $E_G(k)$ across \vdd{the} wavenumber shell $k$, \revi{similarly $\tau_\Omega (k)$ is the local turnover time scale to transfer $\Omega_G(k)$ across $k$}. 
\vdd{Then, by using \eqref{eq:fluxEG}, \eqref{eq:fluxOG} and \eqref{eq:connection} we can obtain the following power laws 
for $\alpha>0$,
	\begin{align}
		& E_G(k) \propto \epsilon^{2/3} k^{(\alpha - 7)/3} \quad
  		\text{for} \; k_{min} \ll k \ll k_{f}, 
		\label{eqn:spec_pred1} \\
		& E_G(k) \propto \xi^{2/3} k^{(-\alpha - 7)/3} \quad
		\text{for} \; k_{f} \ll k \ll k_{max}, 
		\label{eqn:spec_pred2}
	\end{align}
with the two wavenumber regimes denoting the inertial ranges above and below the intermediate forcing wavenumber $k_f$, respectively. Similarly, for $\alpha<0$, the power laws remain unaltered, while their wavenumber range of validity is interchanged between \eqref{eqn:spec_pred1} and \eqref{eqn:spec_pred2}. In a similar fashion, we can derive the power laws for the enstrophy spectra using \eqref{eq:connection}.}
	
	The flux $\Pi$ is a measure of the nonlinear cascade of a conserved quantity in turbulence \citep{AlexakisBiferale2018}. The energy flux \vbj{across} a circle of radius $k$ in the 2D wavenumber space is the total energy transferred from the modes within the circle to the modes outside the circle. Consequently, we define the flux of generalised energy $\Pi_E(k,t)$ and enstrophy $\Pi_\Omega(k,t)$ as 
	\begin{subequations}
		\label{eq:Flux}
		\begin{align}
			\label{eq:EFlux}
			\Pi_E(k,t) &= \sum_{k' \leq k} T_E(k',t), \\
			\label{eq:QFlux}
			\Pi_\Omega(k,t) &= \sum_{k' \leq k} T_\Omega(k',t),
		\end{align}
	\end{subequations}
	where $T_E(k,t)$ and $T_\Omega(k,t)$ are the 
	non-linear generalised energy and enstrophy transfers across $k$ 
	\begin{subequations}
		\label{eq:Transfer}
		\begin{align}
			\label{eq:ETransfer}
			T_E(k,t) &= \sum_{k \leq |{\bf k}| < k + \Delta k} \widehat{\psi}^* ({\bf k}, t) \widehat{\revi{J(q, \psi)}} ({\bf k}, t), \\
			\label{eq:QTransfer}
			T_\Omega(k,t) &= \sum_{k \leq |{\bf k}| < k + \Delta k} \widehat{q}^* ({\bf k}, t) \widehat{\revi{J(q, \psi)}} ( {\bf k}, t),
		\end{align}
	\end{subequations}
	where the sum is performed over the Fourier modes with wavenumber amplitude $k$ in a shell of
	width \final{$\Delta k=1$.} 
	
	In general we cannot determine the direction of cascade of two \vd{quadratic} invariants unless there are special relations between the two like Eq. \eqref{eq:connection}.
	A generalised \cite{Fjortoft1953} argument on the directions of the cascades of two quadratic invariants \vd{$A$ and $B$, with energy spectra $E_A(k)$ and $E_B(k)$, has been formulated by \cite{AlexakisBiferale2022}, which states that if there exists a constant $c > 0$ and an exponent $\beta > 0$, such that $E_A(k)$ and $E_B(k)$ satisfy: 
	\begin{enumerate}
	  \item $|E_A(k)| \leq c k^{-\beta} E_B(k)$, then $A$ cannot cascade forward, or
	  \item $|E_B(k)| \leq c k^{\beta} E_A(k)$, then $B$ cannot cascade inversely.
	\end{enumerate}
\noindent In other words, when $\alpha > 0$ then $E_G$ is transferred towards large scales while $\Omega_G$ is transferred towards small scales. The opposite is true when $\alpha < 0$. In appendix \ref{sec:bounds} we demonstrate this rigorously using mathematical inequalities on the dissipation rates of the generalised energy and enstrophy for positive and negative values of $\alpha$ separately.}

\section{Numerical Results}
\label{sec:results}
\vd{
We numerically integrate Eq. \eqref{eq:qmodel} to \revi{investigate the transition of the cascades, for high Reynolds numbers, by varying} $\alpha$ systematically in the range $-1 \leq \alpha \leq 2$. The domain is taken to be periodic \final{in} both directions \vdd{with dimensions $[0, 2 \pi] \cross [0,2\pi ]$}. The external forcing we \revi{use} to drive the flows is time-independent, monochromatic, and takes the form \revi{$f_q = (-\nabla^2)^{\frac{\alpha}{2}}f_\psi$} with $f_\psi = f_0 \sin(k_f x) \sin(k_f y)$, where $f_0$ denotes the amplitude of the forcing and $k_f$ the forcing wavenumber.  
In our simulations we choose $n=2$ for the hyper-viscous term and $m=2$ for the hypo-viscous term. The details of the numerical method and simulation parameters used are described in Appendix \ref{App:num_setup}. 
}

\subsection{Transition of generalised energy and enstrophy cascades}
%
\vd{
Using the large scale dissipation rates \revi{$\epsilon_- = \nu_- \avg{\psi \nabla^{-4}q}$} and \revi{$\xi_- = \nu_- \avg{q \nabla^{-4}q}$}, we quantify the variation in the generalised energy and enstrophy fluxes to large scales.
} 
For the parameters explored, $\epsilon_{-}$ and $\xi_{-}$ are localised at large length scales giving an estimate of the flux to large scales occurring through an inverse cascade.

\begin{figure}
	\begin{subfigure}[h]{0.49\textwidth}
		\includegraphics[width=\linewidth]{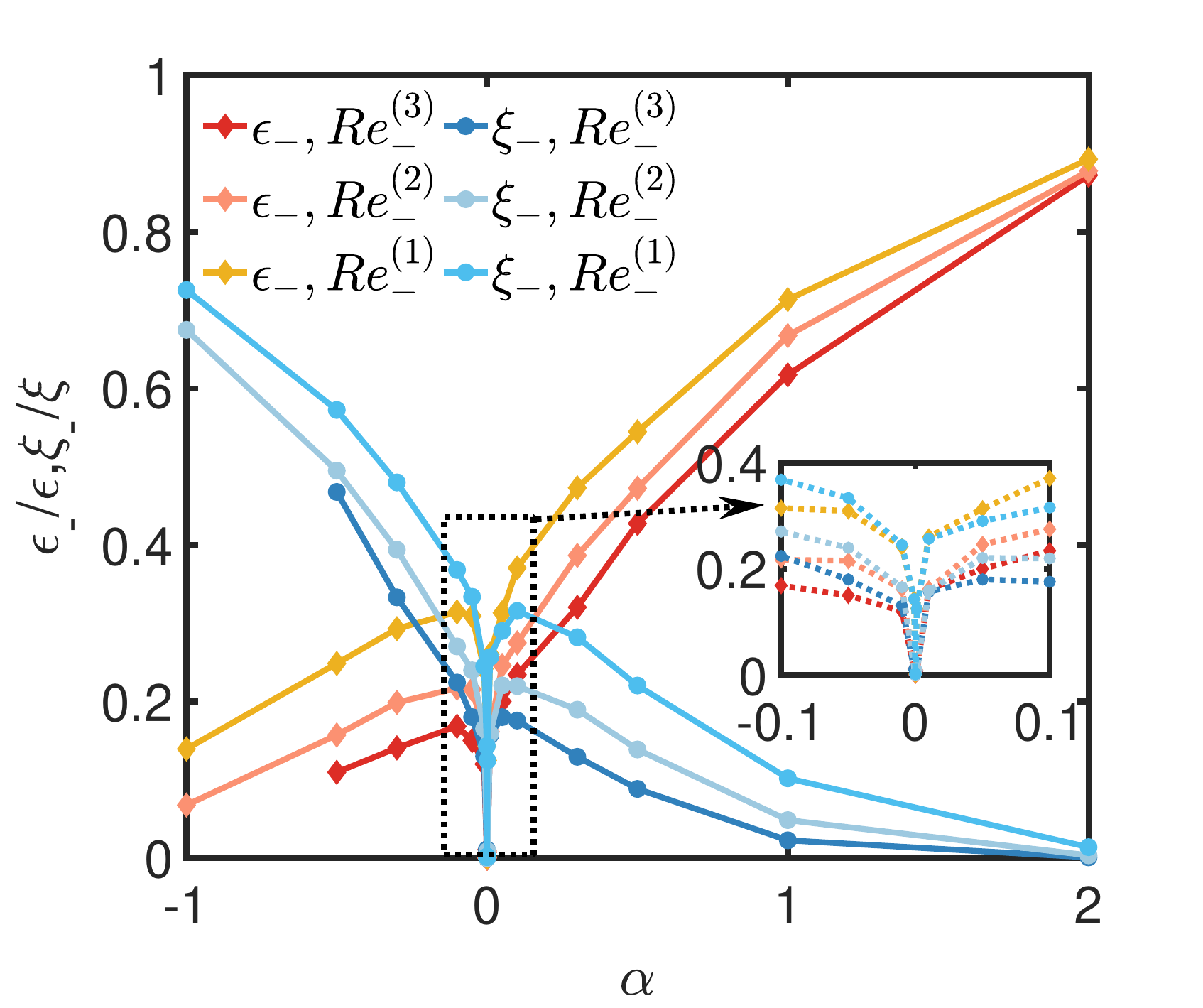}
		\caption{\label{fig:Transition_large}}
	\end{subfigure}
	\begin{subfigure}[h]{0.49\textwidth}
		\includegraphics[width=\linewidth]{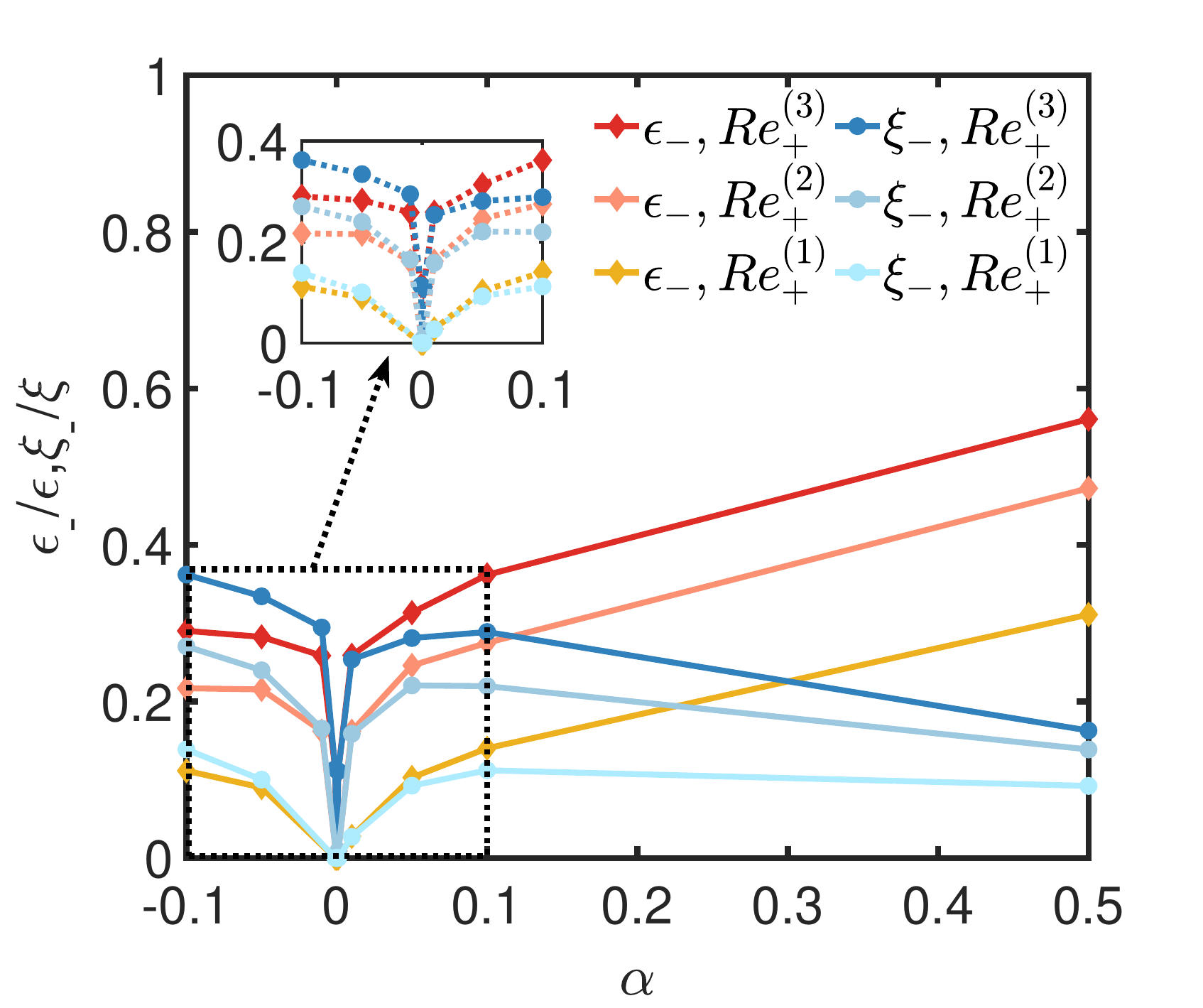}
		\caption{\label{fig:Transition_small}}
	\end{subfigure}
	\caption{\label{fig:transitions} Normalised large scale dissipation rate of generalised energy $\epsilon_{-} / \epsilon$ and enstrophy $\xi_{-} / \xi$ as a function of $\alpha$. Plot a) shows the bifurcation diagram for fixed $Re_+ = 1381$ and increasing values of $\Re_-$, where
		$Re_-^{(1)} = 10^6$, $Re_-^{(2)} = 4 \times 10^7$ and $Re_-^{(3)} = 2 \times 10^9$. Plot b) shows the bifurcation diagram for fixed $Re_- = 4 \times 10^7$ and increasing value of $Re_+$, where $Re_+^{(1)} = 345$, $Re_+^{(2)} = 1381$ and $Re_+^{(3)} = 3450$.}      
\end{figure}

In Fig. \ref{fig:transitions} we show the large scale dissipation rates normalised by their respective injection rates, $\epsilon_{-}/\epsilon$ and $\xi_{-}/\xi$, as a function of $\alpha$. \vdd{The normalised large and small scale quantities are related by the expression $\frac{\epsilon_-}{\epsilon}=1-\frac{\epsilon_+}{\epsilon}$ from 
\eqref{eq:energybal} and 
$\frac{\xi_-}{\xi}=1-\frac{\xi_+}{\xi}$ from 
\eqref{eq:enstrophybal}.} 
In Fig. \ref{fig:Transition_large} we consider different values of $Re_{-}$ with fixed $Re_+$, while in Fig. \ref{fig:Transition_small} we consider different values of $Re_{+}$ with fixed $Re_-$. We find that for $\alpha = 2$ the energy dissipation is predominantly at large length scales due to the inverse cascade of energy in 2D Navier-Stokes turbulence. On the other hand, for $\alpha = -1$ we observe the transition of the cascades with the energy to be predominantly dissipated at small scales suggesting a forward cascade of energy. For intermediate values of $\alpha$ we find that energy is dissipated at both large and small scales implying the presence of a \vd{bidirectional cascade.}

As seen from Fig. \ref{fig:Transition_large}, increasing $Re_-$ for a given $Re_+$ 
leads to \final{weaker inverse cascades} of both the generalised energy and enstrophy. 
In Figure \ref{fig:Transition_small} increasing $Re_+$ for a fixed $Re_-$ leads to smaller forward cascade of both the generalised energy and enstrophy. 
So, the cascades of generalised energy and enstrophy depend on the values of $Re_+$ and $Re_-$ we have considered. \vdd{
Further studies focusing on particular values of $\alpha$ and systematically increasing $Re_+$ and $Re_-$ could shed light on whether the dissipation rates become finite or not in the limit of infinite Reynolds.}

\vd{Close to $\alpha = 0$ we find that the behaviour depends} strongly on the control parameters $Re_+$ and $Re_-$ as the nonlinearity, which leads to the cascade, \vd{diminishes} as $\alpha \rightarrow 0$.
At $\alpha = 0$ there is no cascade as the nonlinearity vanishes, i.e. $J(\psi,\psi) = 0$ by definition. \revi{Hence,} both the forcing and dissipation \revi{are} localised at the wavenumber $k_f$. \nnew{For low enough values of \final{$|\alpha|$}, temporal chaotic behaviour is observed. However, based on the energy spectrum we find turbulent flows for $\alpha\gtrsim 0.001$ and $\alpha\lesssim -0.001$ in our numerical simulations.} We refer to $\alpha = 0$ as the threshold where the cascades vanish for any value of $Re_+$ and $Re_-$. For the simulation parameters that are explored here, the small scale dissipation dominates over the large scale dissipation near $\alpha \approx 0$ implying that the dissipation due to large scale friction is negligible, thus \vdd{$\epsilon_- / \epsilon \ll 1$}.
The exact behaviour close to 
$\alpha = 0$ will be discussed in section \ref{sec:near_threshold}. 

\begin{figure}
	\centering \includegraphics[width=0.5\linewidth]{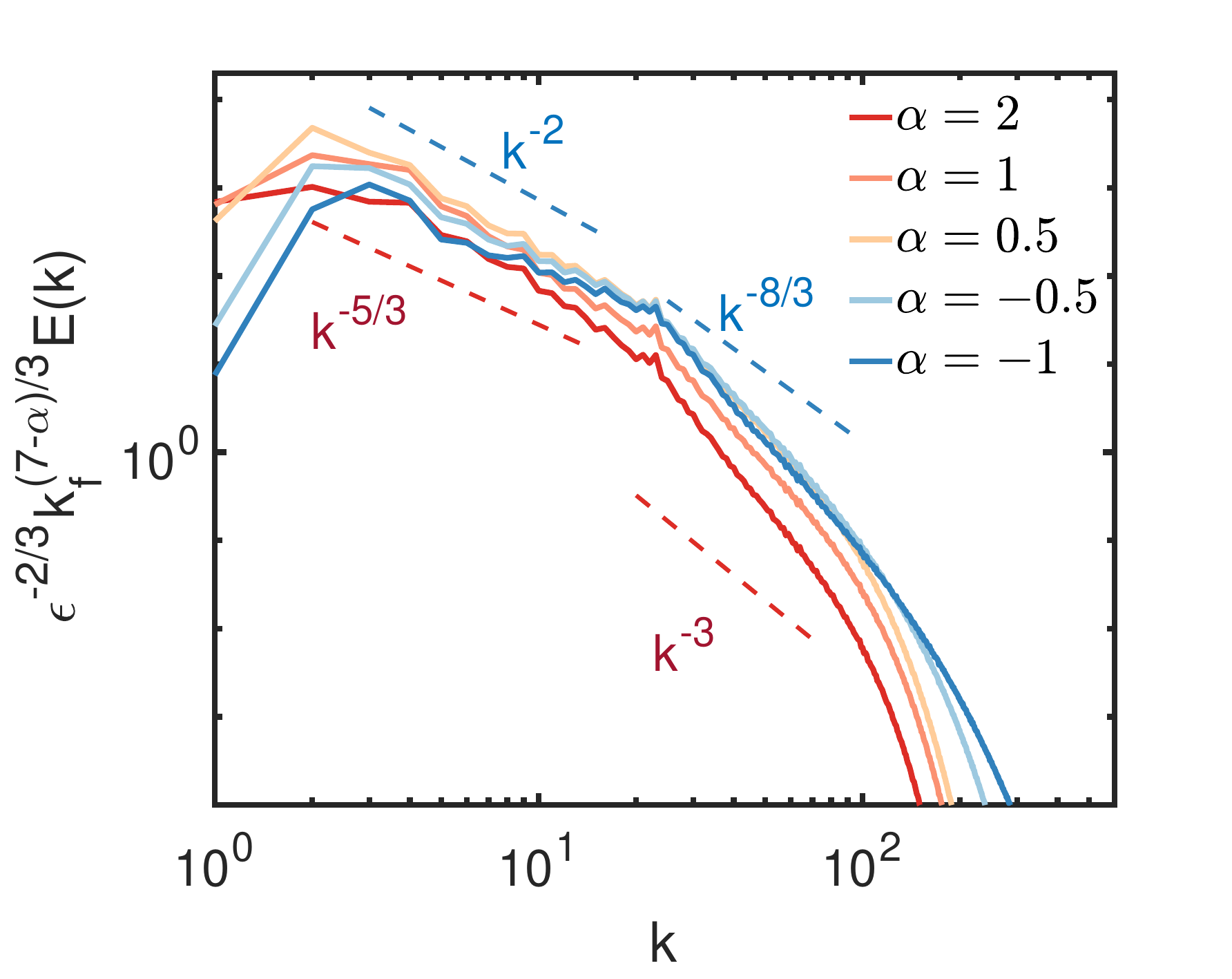}
	\caption{\label{fig:Spectra} Energy spectra for different $\alpha$ along with their theoretical exponents in red ($\alpha=2$) and blue ($\alpha=-1$) are shown as dashed line. The spectra are obtained from DNS with $Re_+=1381$, $Re_-=4\times10^7$ and $k_f=16\sqrt{2}$.}
\end{figure}

In Fig. \ref{fig:Spectra} we show the generalised energy spectrum for different values of $\alpha$. The dashed lines show the phenomenological predictions from Eqs. \eqref{eqn:spec_pred1}, \eqref{eqn:spec_pred2}. The red dashed lines show the predictions for $\alpha = 2$ and the blue dashed lines show the predictions for $\alpha = -1$. We see that the generalised energy spectra in the range $k_{min} < k < k_f$ have exponents close to the predictions \eqref{eqn:spec_pred1}, while in the range $k_f < k < k_{max}$ the exponents are far from the phenomenological predictions. \vd{This discrepancy might arise out of intermittency effects \citep{ishiharaetal2009}. In particular, for the 2D Navier-Stokes turbulence ($\alpha = 2$), \cite{boffetta2010evidence} report that the exponent in the range $k_f < k < k_{max}$ depends on the small scale Reynolds number. Simulations at higher resolutions} are required to shed light on the differences 
\vdd{between the spectral exponents from DNS for the different values of $\alpha$ and the 
Kolmogorov-like scalings from \eqref{eqn:spec_pred1}, \eqref{eqn:spec_pred2}.} 

\begin{figure}
	\begin{subfigure}[h]{0.49\textwidth}
		\includegraphics[width=\linewidth]{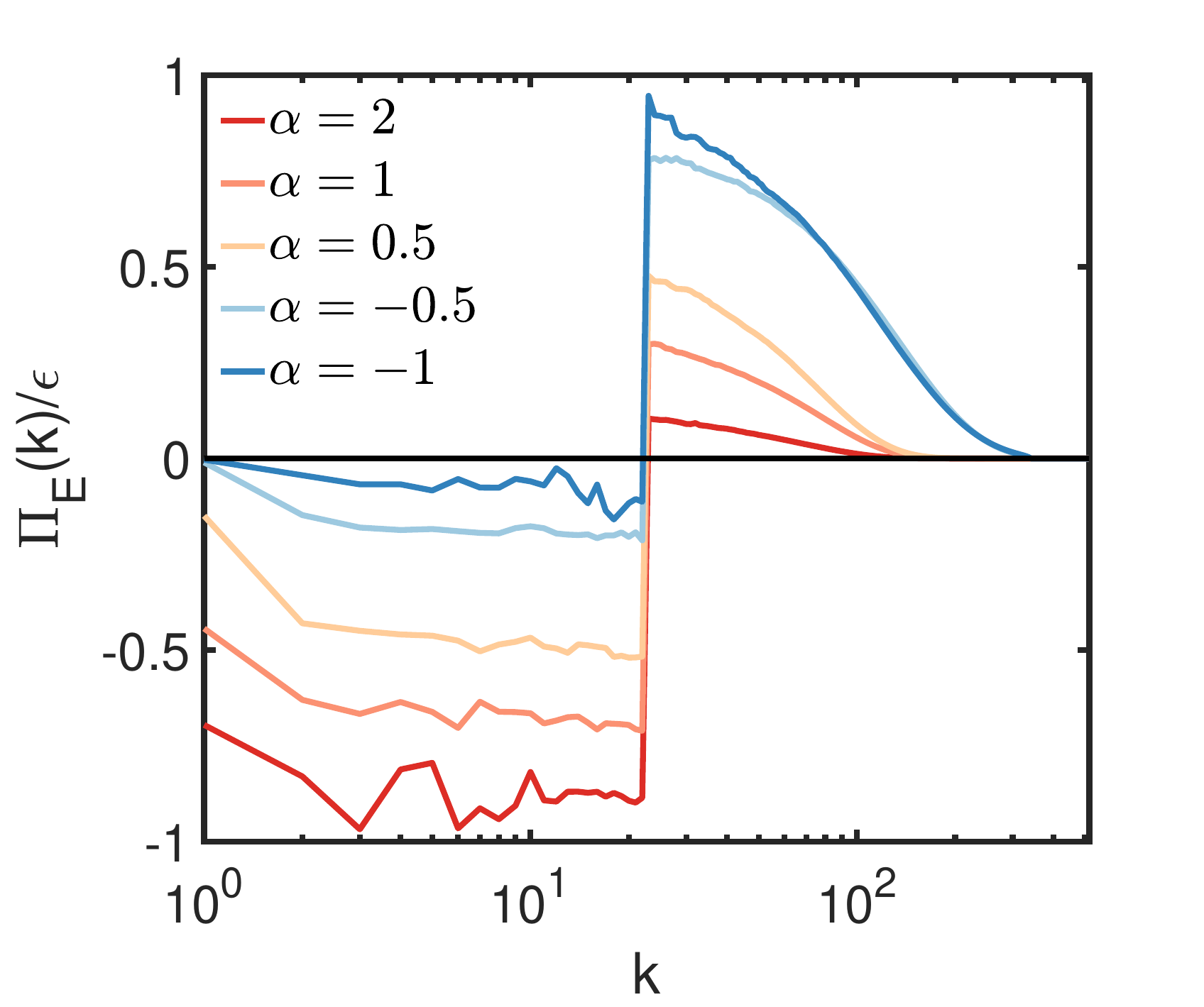}
		\caption{\label{fig:Flux_PiEk}}
	\end{subfigure}
	\begin{subfigure}[h]{0.49\textwidth}
		\includegraphics[width=\linewidth]{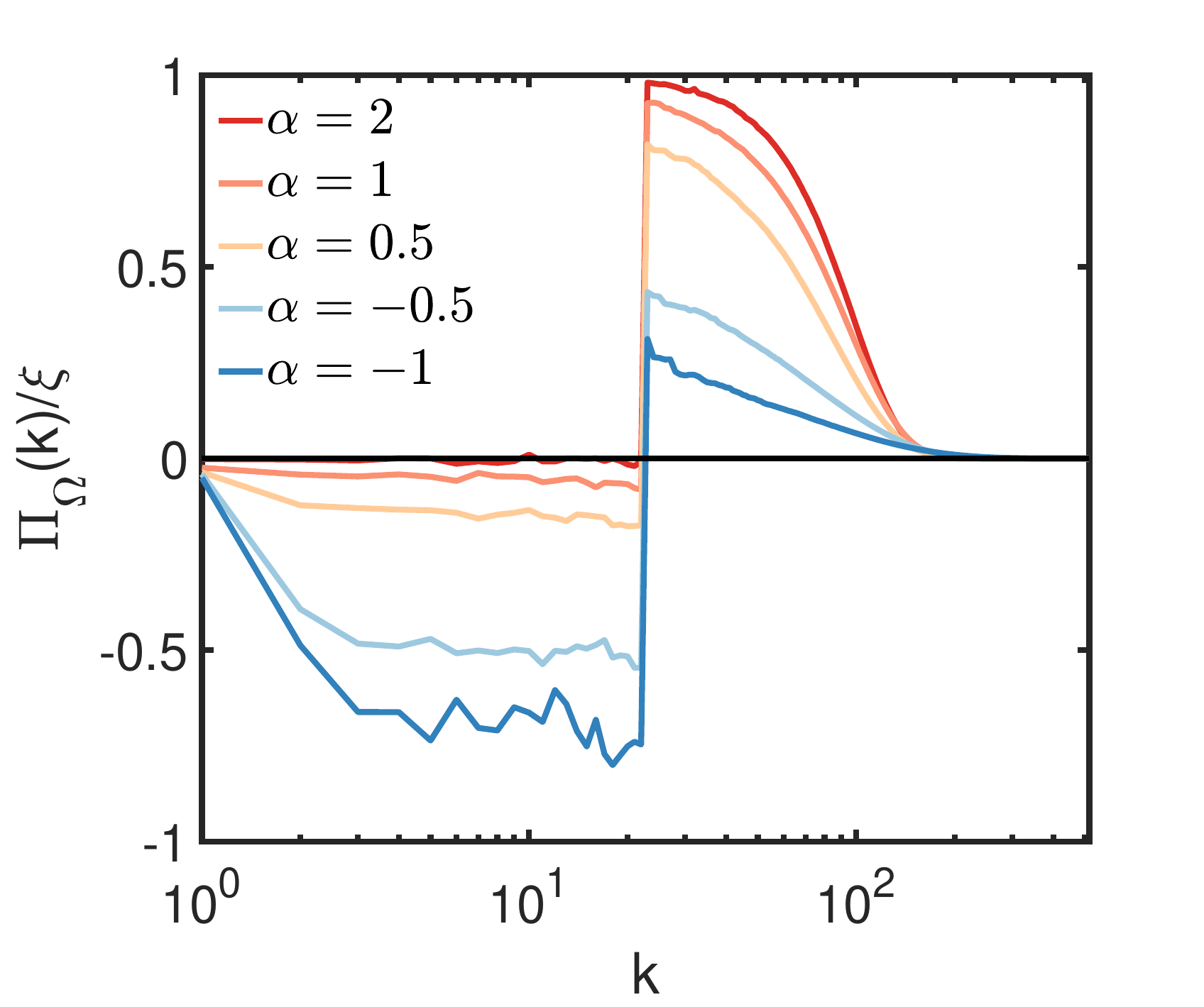}
		\caption{\label{fig:Flux_PiOk}}
	\end{subfigure}
	\caption{\label{fig:fluxes} a) Normalised energy flux and b) normalised enstrophy flux for different $\alpha$. The fluxes are obtained from DNS with $Re_+=1381$, $Re_-=4\times10^7$ and $k_f=16\sqrt{2}$.}
\end{figure}

Figure \ref{fig:fluxes} shows the fluxes of generalised energy and enstrophy, normalised with their respective injection rates, for different values of $\alpha$. For the Navier-Stokes model ($\alpha = 2$) we see that $E_G$ cascades mostly to large scales (or low $k$)\vbj{,}  while $\Omega_G$ cascades mostly to small scales (or high $k$). For the $\alpha = -1$ we find the opposite scenario where $E_G$ mostly cascades to small scales while $\Omega_G$ cascades to large scales. 
\vd{
Intermediate values of $\alpha$ show bidirectional cascades, where $E_G$ and $\Omega_G$ cascade to both large and small scales. This \vdd{transition of cascades} is similar to those observed in other turbulent systems \citep{celani2010turbulence,Seshasayananetal2014,benavides2017critical,van2020critical}. 
}

%

\vd{
Generally, a bidirectional cascade can be observed if the dynamical properties of the flow at wavenumbers $k \ll k_f$ can support an inverse transfer, and at the same time they can support a forward transfer at wavenumber $k \gg k_f$ for a range of values of the parameters \citep{AlexakisBiferale2018}. 
In generalised 2D turbulence, the inverse transfer can be attributed to instabilities that couple the forced modes to large-scale 2D modes, transferring energy to larger scales, which is typical for the $\alpha = 2$ model \citep{alexakis2018three}. For the $\alpha = 1$, the forward transfer \vdd{can be} attributed to the filament instabilities, which generate smaller scales \citep{Heldetal1995,ScottDritchel2014}. A mechanism to inhibit the instability of a filament is the presence of a background strain due to a distant \vdd{vortex}. However, it has been found that the region of influence of the vortices decreases as we reduce $\alpha$ \citep[see][]{IwayamaWatanabe2010,Cartonetal2016}. This can explain why the forward flux of the generalised energy increases relative to the inverse flux as $\alpha$ decreases.
}

\begin{figure}
	\centering
	\begin{subfigure}[h]{0.4\textwidth}
		\includegraphics[width=\linewidth]{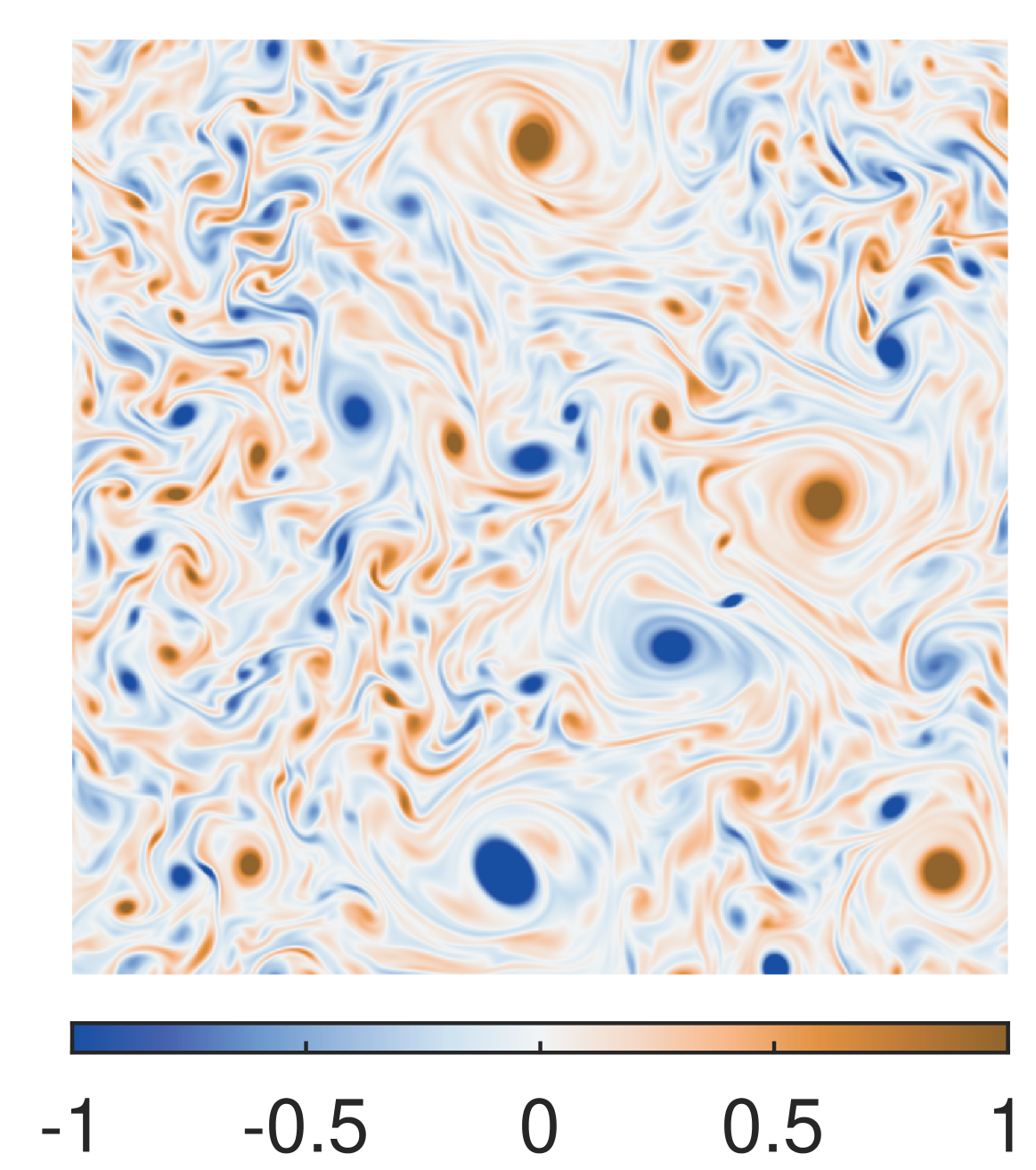}
		\caption{\label{fig:gen_vort_1}}
	\end{subfigure}
	\hspace{9pt}
	\begin{subfigure}[h]{0.4\textwidth}
		\includegraphics[width=\linewidth]{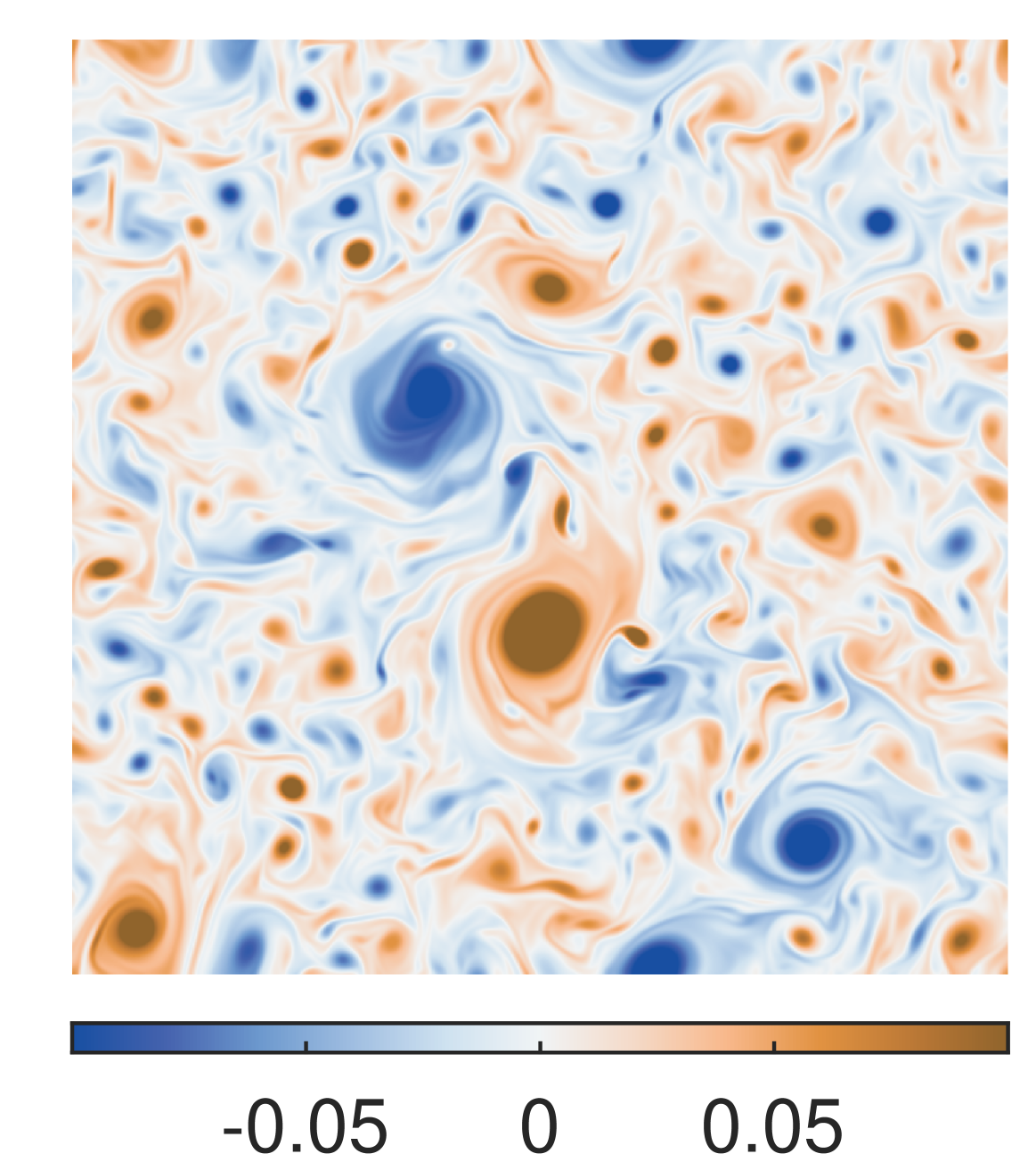}
		\caption{\label{fig:gen_vort_2}}
	\end{subfigure}
	\begin{subfigure}[h]{0.4\textwidth}
		\includegraphics[clip,width=\linewidth]{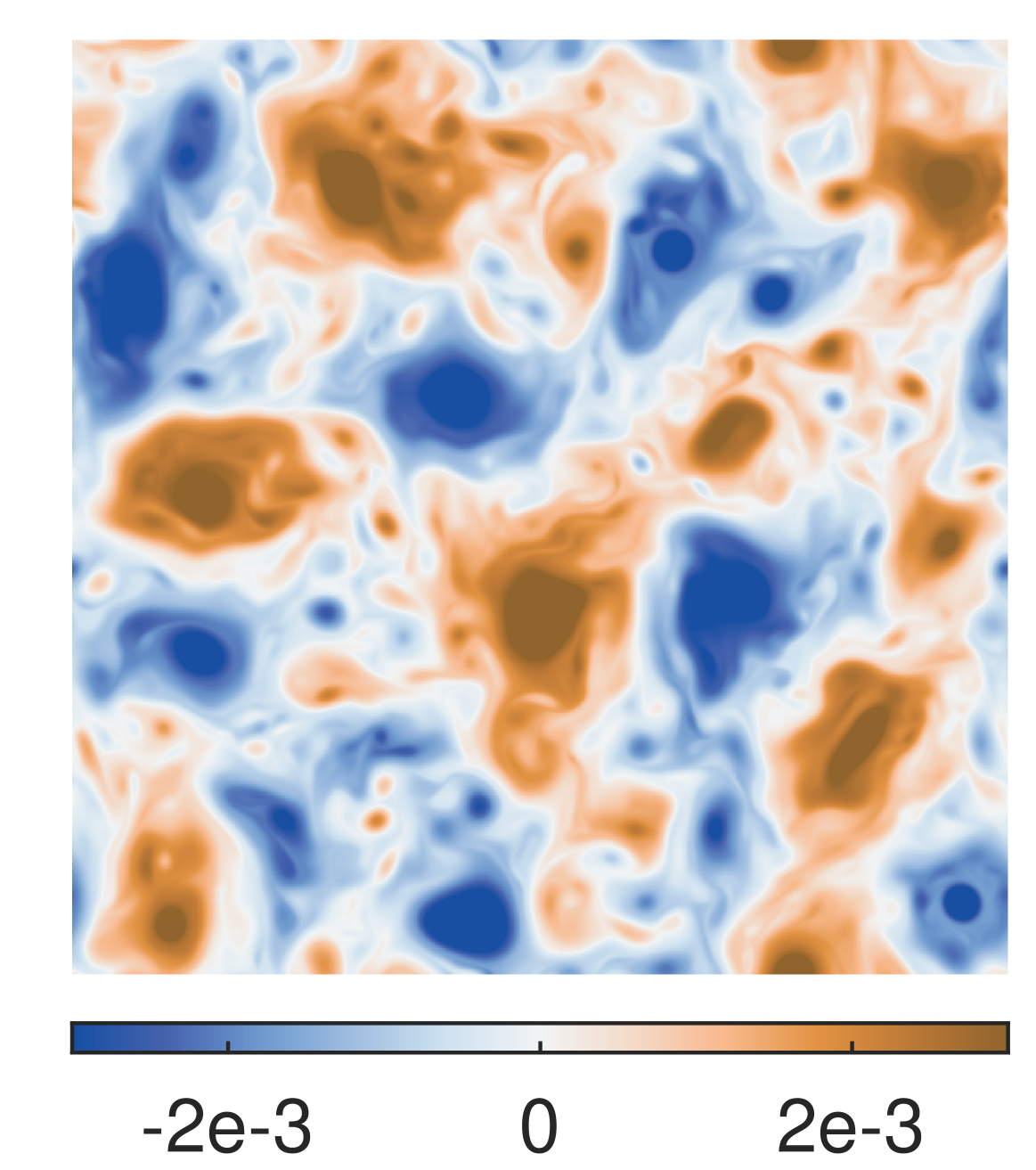}
		\caption{\label{fig:gen_vort_3}}
	\end{subfigure}
	\hspace{0.1pt}
	\begin{subfigure}[h]{0.4\textwidth}
		\includegraphics[clip,width=\linewidth]{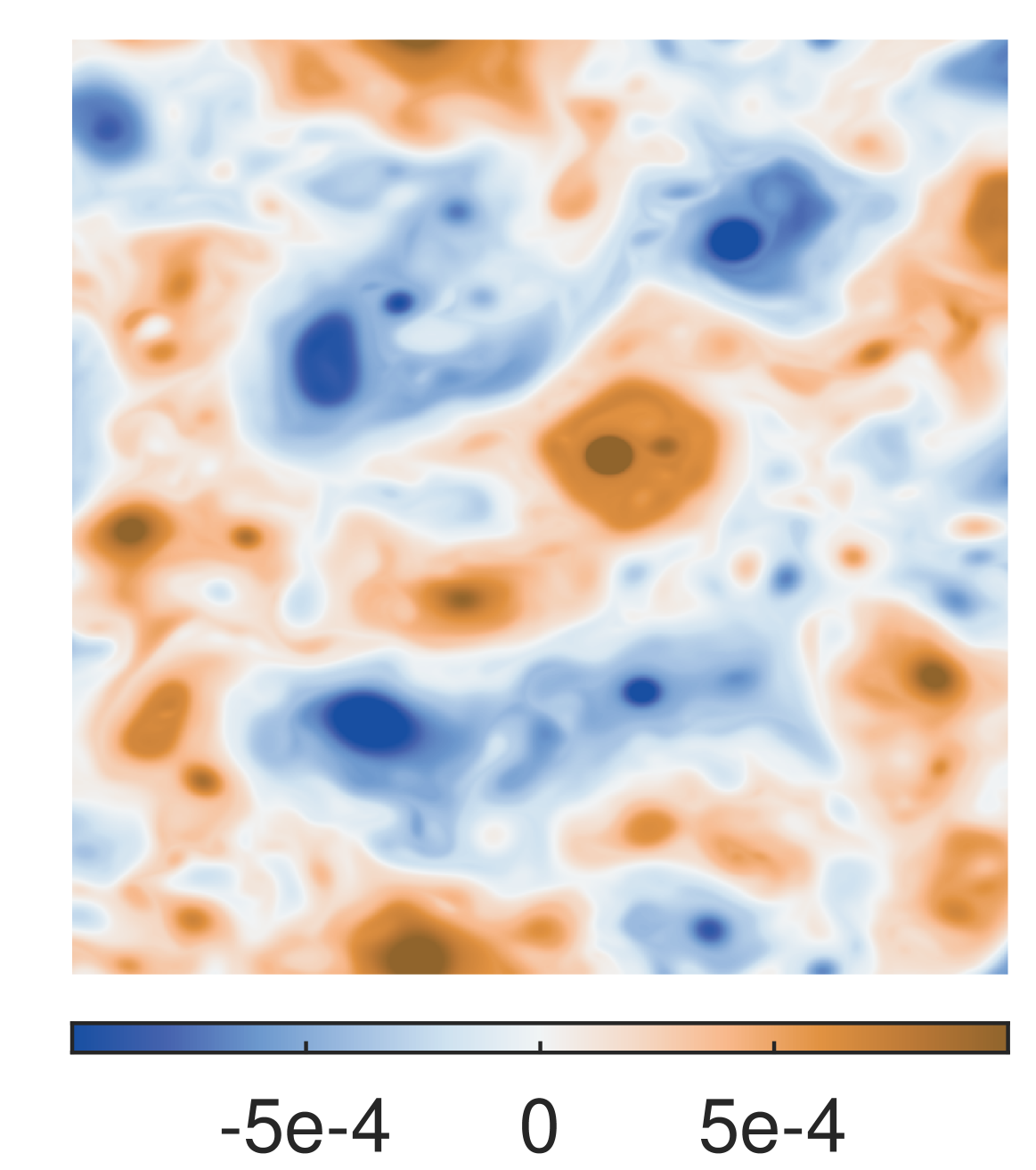}
		\caption{\label{fig:gen_vort_4}}
	\end{subfigure}
	\caption{\label{fig:generalised_fields} 
	\vd{Generalised vorticity field for a) $\alpha=2.0$ b) $\alpha=1.0$ c) $\alpha=-0.5$ d) $\alpha=-1.0$. For $\alpha=2,1$ and $-0.5$ the colorbar has been capped at $50\%$ of the maximum value and at $10\%$ for $\alpha=-1$ to emphasise the structures of the fields. The fields are obtained from DNS with $Re_+=1381$, $Re_-=4 \times 10^7$ and $k_f=16\sqrt{2}$.}
	}
\end{figure}

\begin{figure}
	\centering
	\begin{subfigure}[h]{0.4\textwidth}
		\includegraphics[width=\linewidth]{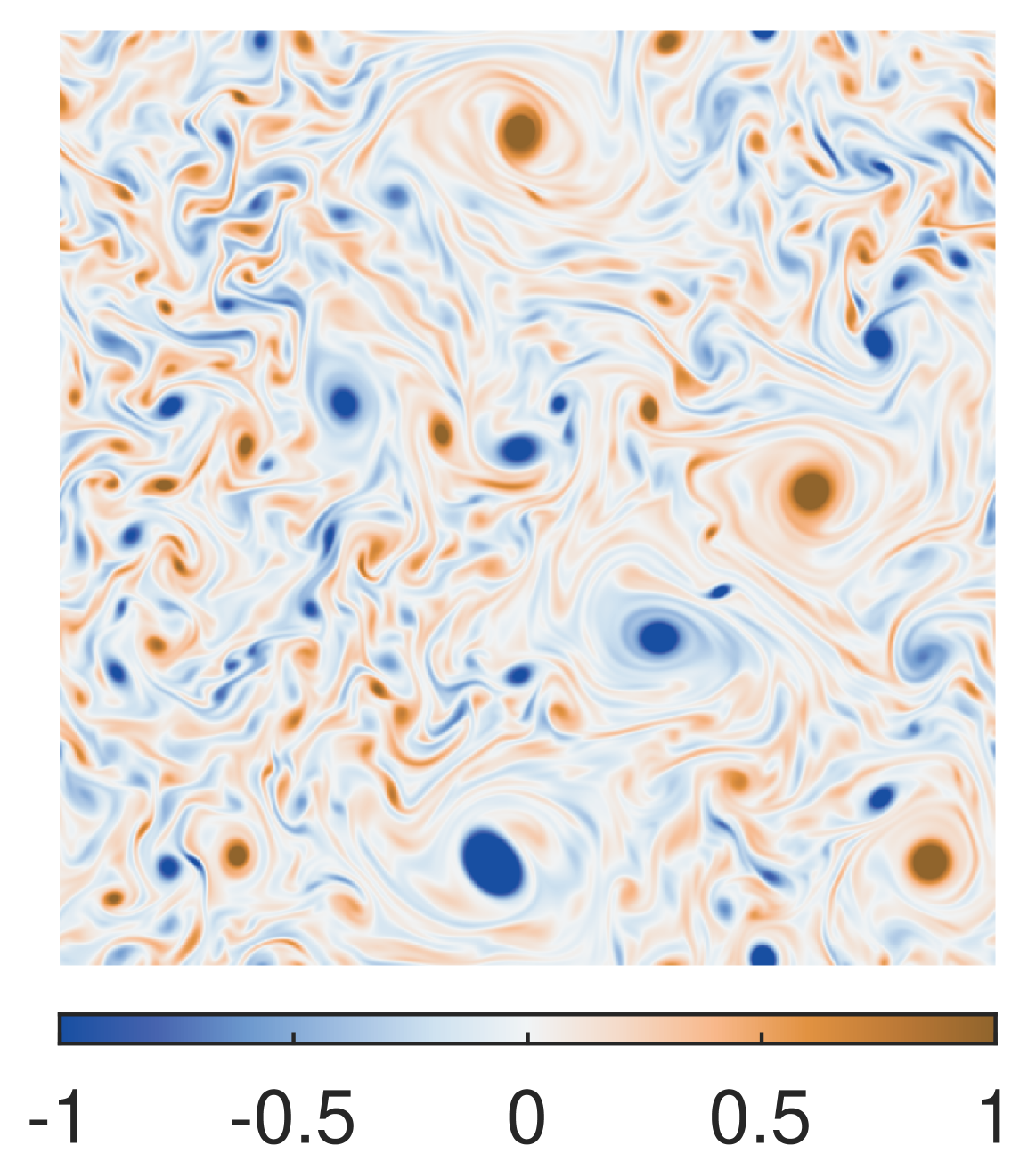}
		\caption{\label{fig:cla_vort_1}}
	\end{subfigure}
	\hspace{3pt}
	\begin{subfigure}[h]{0.4\textwidth}
		\includegraphics[width=\linewidth]{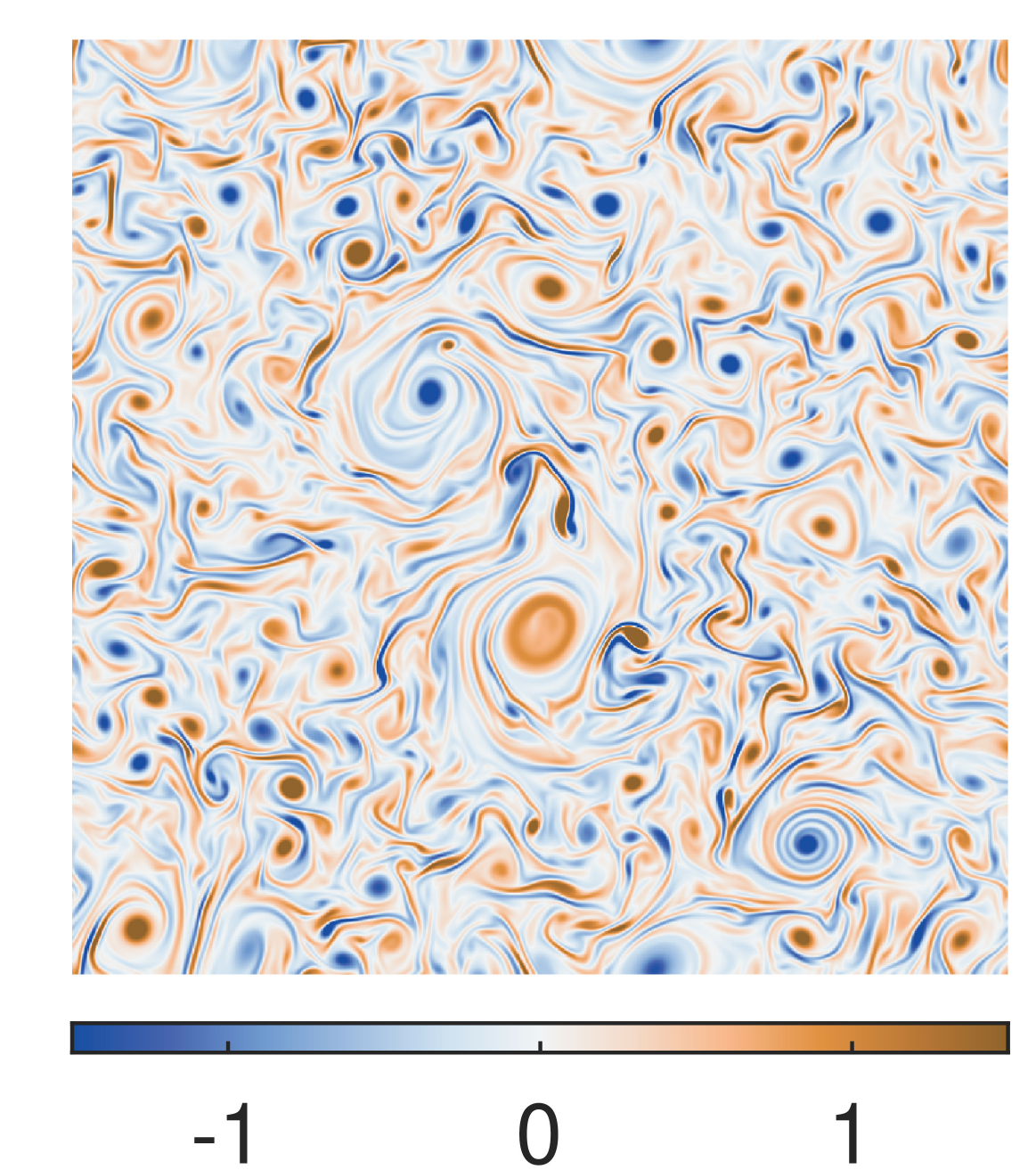}
		\caption{\label{fig:cla_vort_2}}
	\end{subfigure}
	\begin{subfigure}[h]{0.4\textwidth}
		\includegraphics[width=\linewidth]{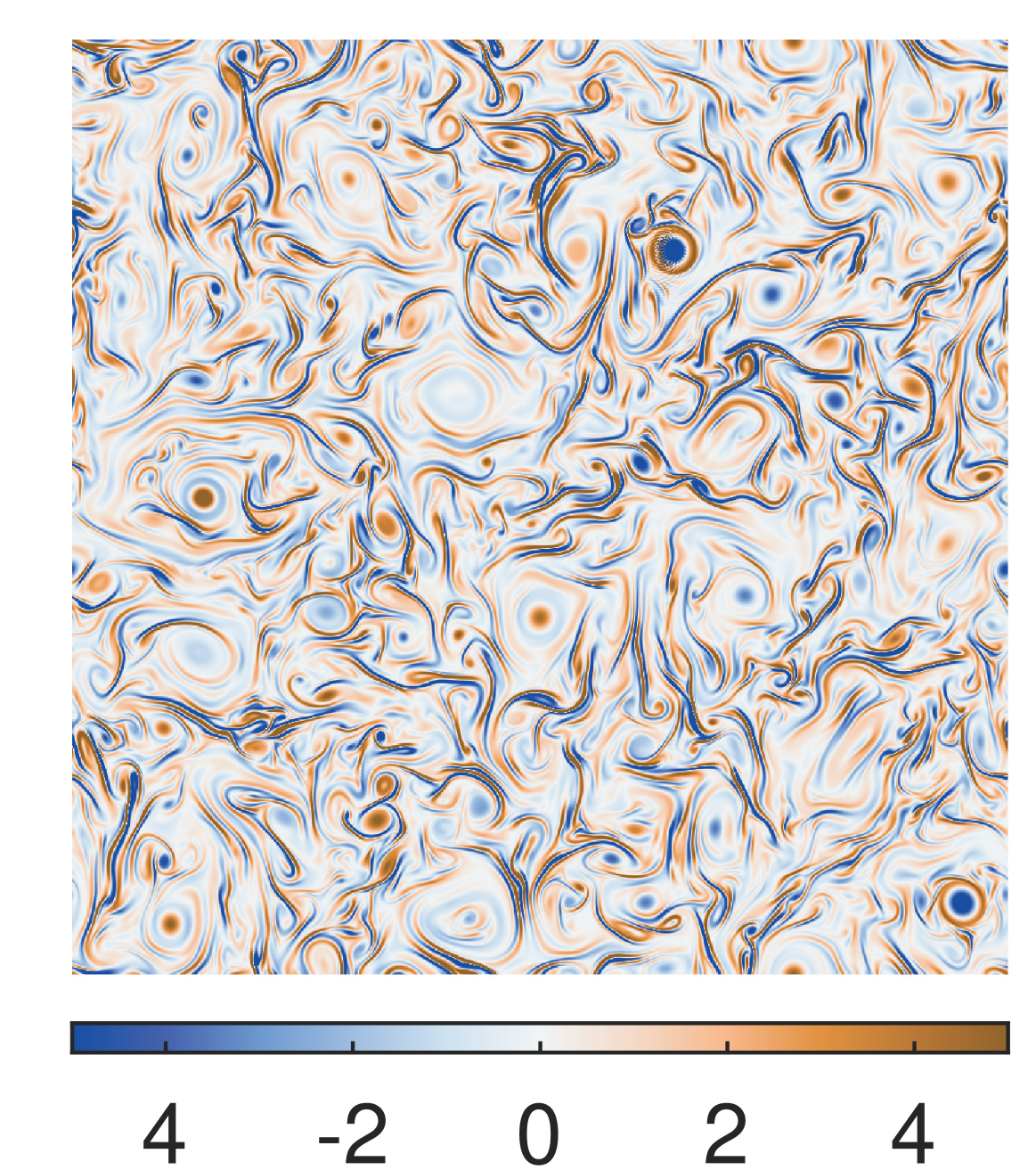}
		\caption{\label{fig:cla_vort_3}}
	\end{subfigure}
	\hspace{0.1pt}
	\begin{subfigure}[h]{0.4\textwidth}
		\includegraphics[width=\linewidth]{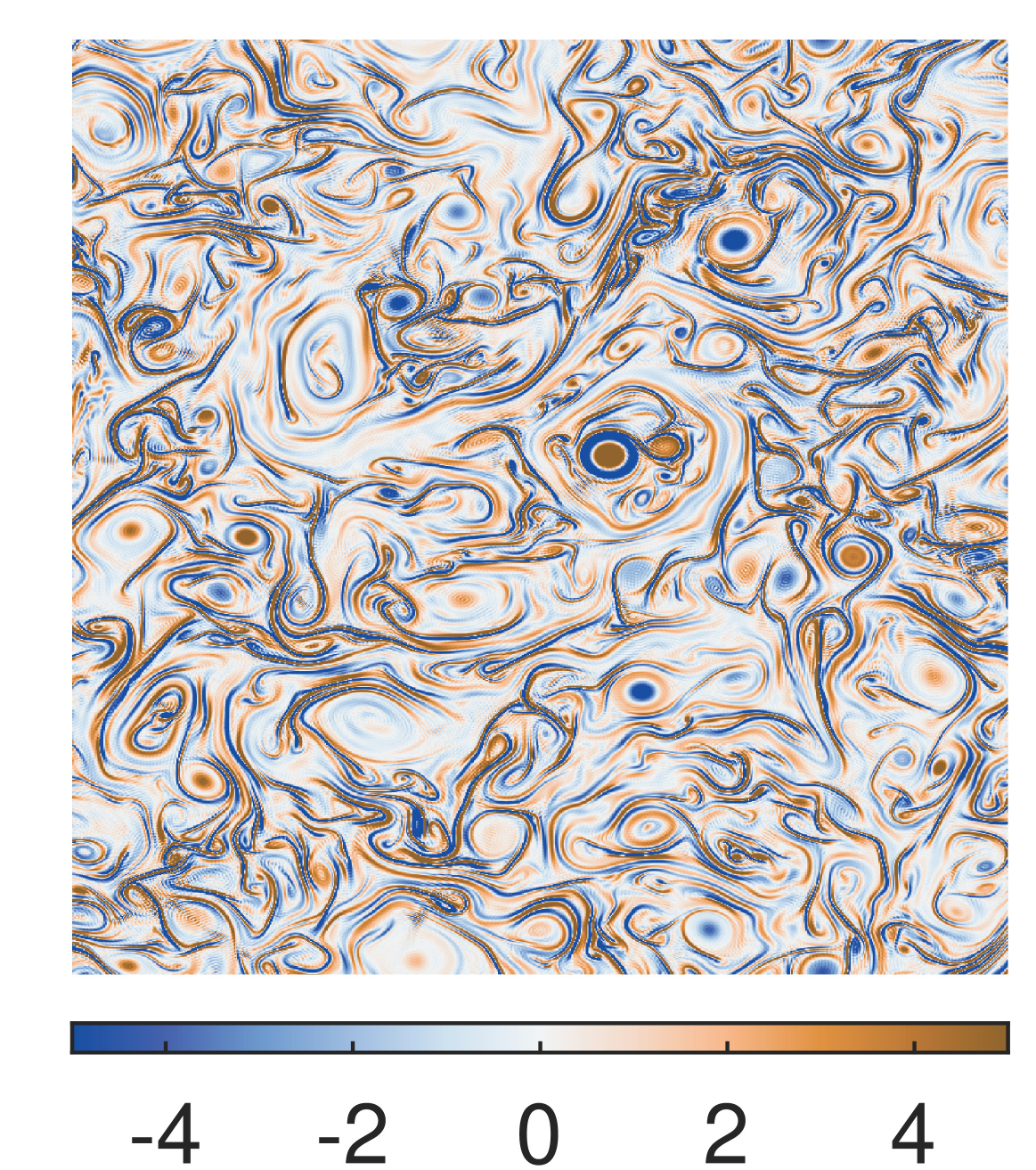}
		\caption{\label{fig:cla_vort_4}}
	\end{subfigure}
	\caption{\label{fig:classical_fields} 
	\vd{Classical vorticity field for a) $\alpha=2.0$ b) $\alpha=1.0$ c) $\alpha=-0.5$ d) $\alpha=-1.0$. For $\alpha=2,1$ and $-0.5$ the colorbar has been capped at $50\%$ of the maximum value and at $10\%$ for $\alpha=-1$ to emphasise the structures of the fields. The fields are obtained from DNS with $Re_+=1381$, $Re_-=4 \times 10^7$ and $k_f=16\sqrt{2}$.}
	}
\end{figure}

\vd{
Figure \ref{fig:generalised_fields} shows the generalised vorticity field for different values of $\alpha$. For $\alpha = 2$, we get the array of vortices that are typically observed in 2D Navier-Stokes,} and are generated by the inverse cascade of energy. As $\alpha$ is reduced we see that the vortices become diffused, \vdd{with their sizes becoming larger}. For $\alpha= 1$, \cite{Cartonetal2016} found that the distance of vortex merging was generally smaller than for $\alpha=2$. It was explained that for a point vortex of the form $q(r) = \delta(r)$, \vdd{where $\delta(r)$ is the Dirac delta function, the velocity field ${\bf u} \propto 1/r^2$ for the SQG model ($\alpha = 1$), while ${\bf u} \propto 1/r$ for the Navier-Stokes model ($\alpha = 2$) from the core of the vortex}. \revi{Hence, the region of influence of the vortices decreases as we reduce $\alpha$, which could explain the reduction in both vortex merging and inverse transfer of energy.} 
\vd{
To distinguish the behaviour of the generalised \vdd{vorticity} from the classical vorticity, we show in Fig. \ref{fig:classical_fields} the contour plots of the classical vorticity} 
defined as \revi{$\omega = -\nabla^2 \psi$} for all values of $\alpha$.
As seen from the plots the classical vorticity goes to smaller length scales as $\alpha$ is reduced. \vd{For $\alpha < 0$ we observe a few vortices along with many filamentary structures, due to the forward cascade of generalised energy, that resemble the thin vortex filaments observed in 3D Navier-Stokes turbulence \citep{ishiharaetal2009}}. 

\subsection{Near $\alpha = 0$} \label{sec:near_threshold}

We investigate the behaviour of the system close to $\alpha = 0$. At $\alpha = 0$ the nonlinearity vanishes and the flow becomes laminar. As one moves away from the threshold, the flow remains laminar until it undergoes a linear instability. 
\vd{
By numerically solving the eigenvalue problem of the linearised equations at small values of $\alpha$, we \revi{find} the primary instability to occur at length scales smaller than the laminar flow, unlike the negative viscosity instability that is observed in many 2D flows \citep{SivashinskyYakhot1985}. This suggests that there is no inverse transfer of energy to scales larger than the forcing length scale for sufficiently small values of $\alpha$. Inverse transfer of energy can be excited by secondary instabilities, as we move away from the primary instability threshold.
The threshold of the primary instability is found to be the same for both positive and negative values of $\alpha$. For \vdd{convenience} we report on the critical point of the instability $\alpha_c$ as we increase $\alpha$ away from zero to positive values. 
}

\begin{figure}
	\begin{subfigure}[h]{0.49\textwidth}
		\includegraphics[width=\linewidth]{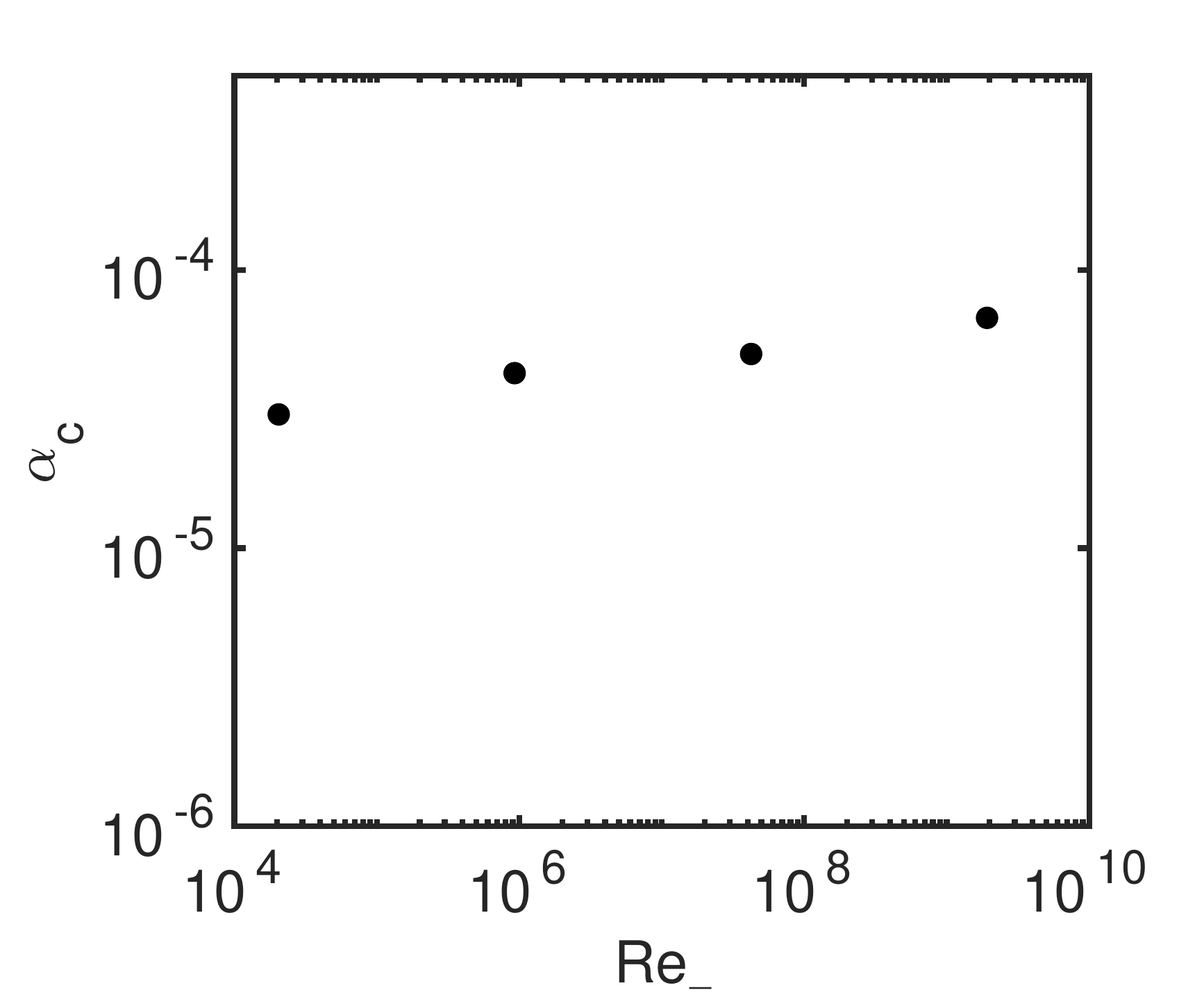}
		\caption{\label{fig:threshold_Re-}}
	\end{subfigure}
	\begin{subfigure}[h]{0.49\textwidth}
		\includegraphics[width=\linewidth]{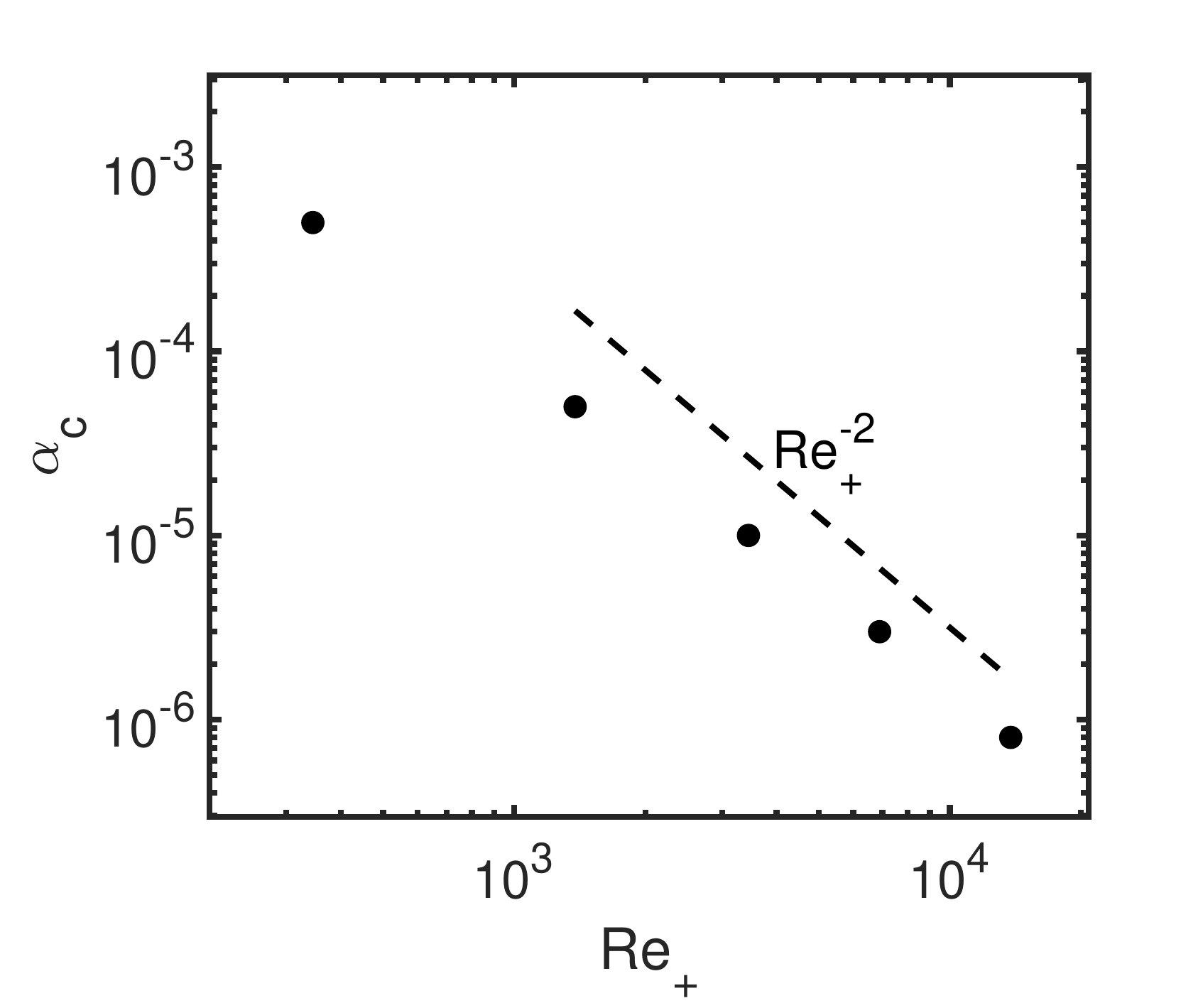}
		\caption{\label{fig:threshold_Re+}}
	\end{subfigure}
	\caption{\label{fig:instability_threshold} Figure shows the variation of $\alpha_c$ as a function of a) $\Re_{-}$ and b) $\Re_{+}$. The dashed line in figure b) denotes the scaling $\alpha_c \propto \Re^{-2}_+$.}
\end{figure}

In Fig. \ref{fig:instability_threshold} we show the critical point of the instability $\alpha_c$ as a function of $Re_{-}$ and $Re_{+}$. For the parameters of $Re_{+} = 1381 $, $Re_{-} = 4\times 10^7 $ at $\alpha = 0$ the large scale dissipation is negligible, thus we find that the instability threshold $\alpha_c$ is almost independent of $Re_{-}$ (see Fig. \ref{fig:threshold_Re-}). We also find that the threshold decreases like a power law as $Re_+$ increases (see Fig. \ref{fig:threshold_Re+}), thus neglecting the effect of large scale \vd{dissipation} we expect that $\alpha_c \rightarrow 0$ as $Re_+ \rightarrow \infty$. The dependence on $Re_+$ indicates that the behaviour very close to the critical point is sensitive to the \vdd{hyper-viscous coefficient.} 

\vd{
The scaling law of $\alpha_c$ with $Re_+$ we observe in Fig.\ref{fig:instability_threshold} can be derived by doing a perturbation around $\alpha = 0$. Details of the perturbation analysis can be found in appendix \ref{sec:scaling_law}. In essence, we perturb the laminar state and then we balance the non-linear term with the small scale dissipative term to obtain the following scaling law for the threshold of the instability 
\begin{equation}
	\alpha_c \propto Re_{+}^{-2}.
\end{equation} 
This scaling law is denoted by the dashed line in Fig. \ref{fig:threshold_Re+} and has a good agreement with the thresholds found numerically for different $Re_+$. 
}

\subsection{Locality of cascades - $\alpha$ dependence}
\begin{figure}
	\begin{subfigure}[h]{0.24\textwidth}
		\includegraphics[width=\linewidth]{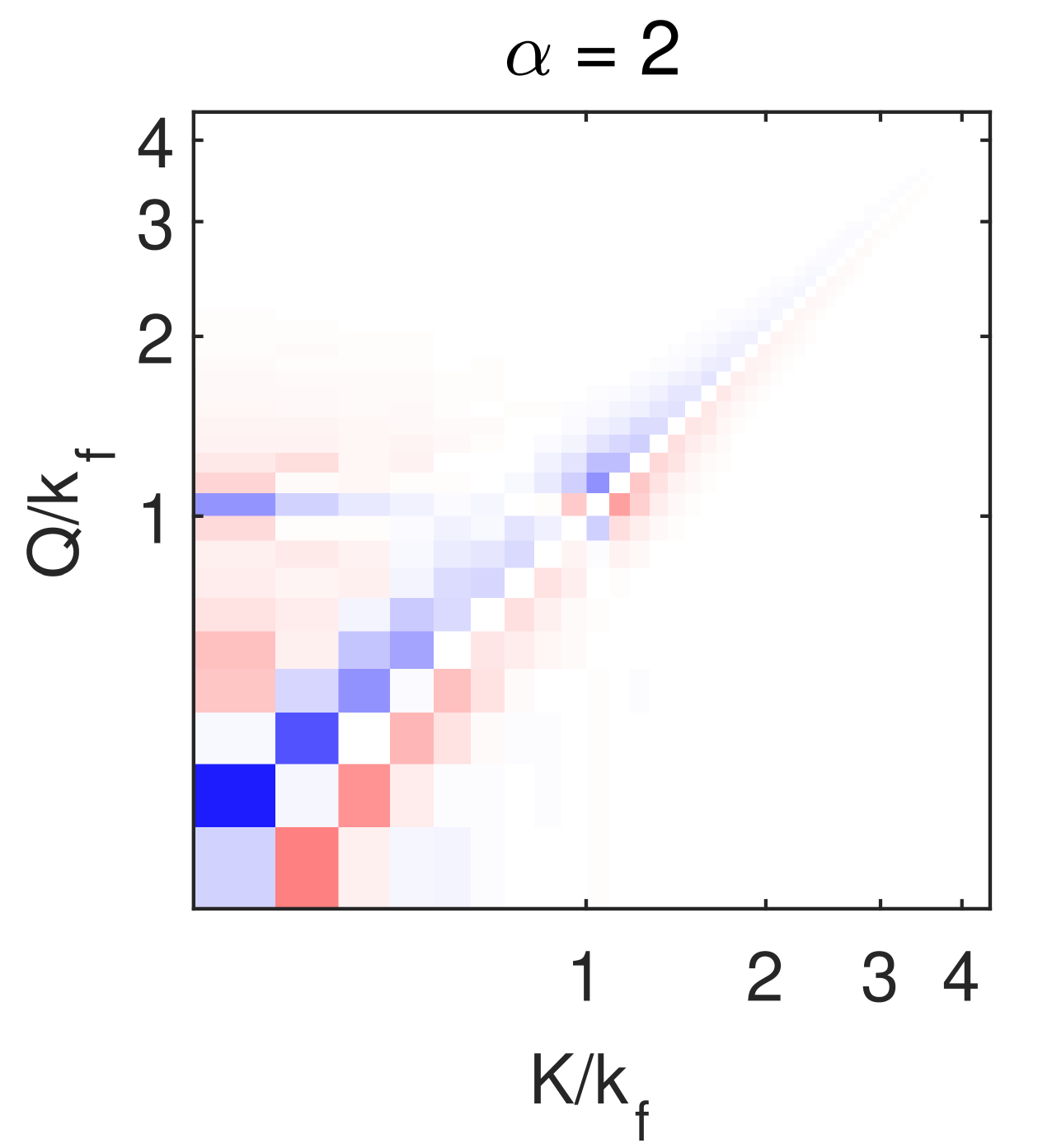}
		\caption{}
	\end{subfigure}
	\begin{subfigure}[h]{0.24\textwidth}
		\includegraphics[width=\linewidth]{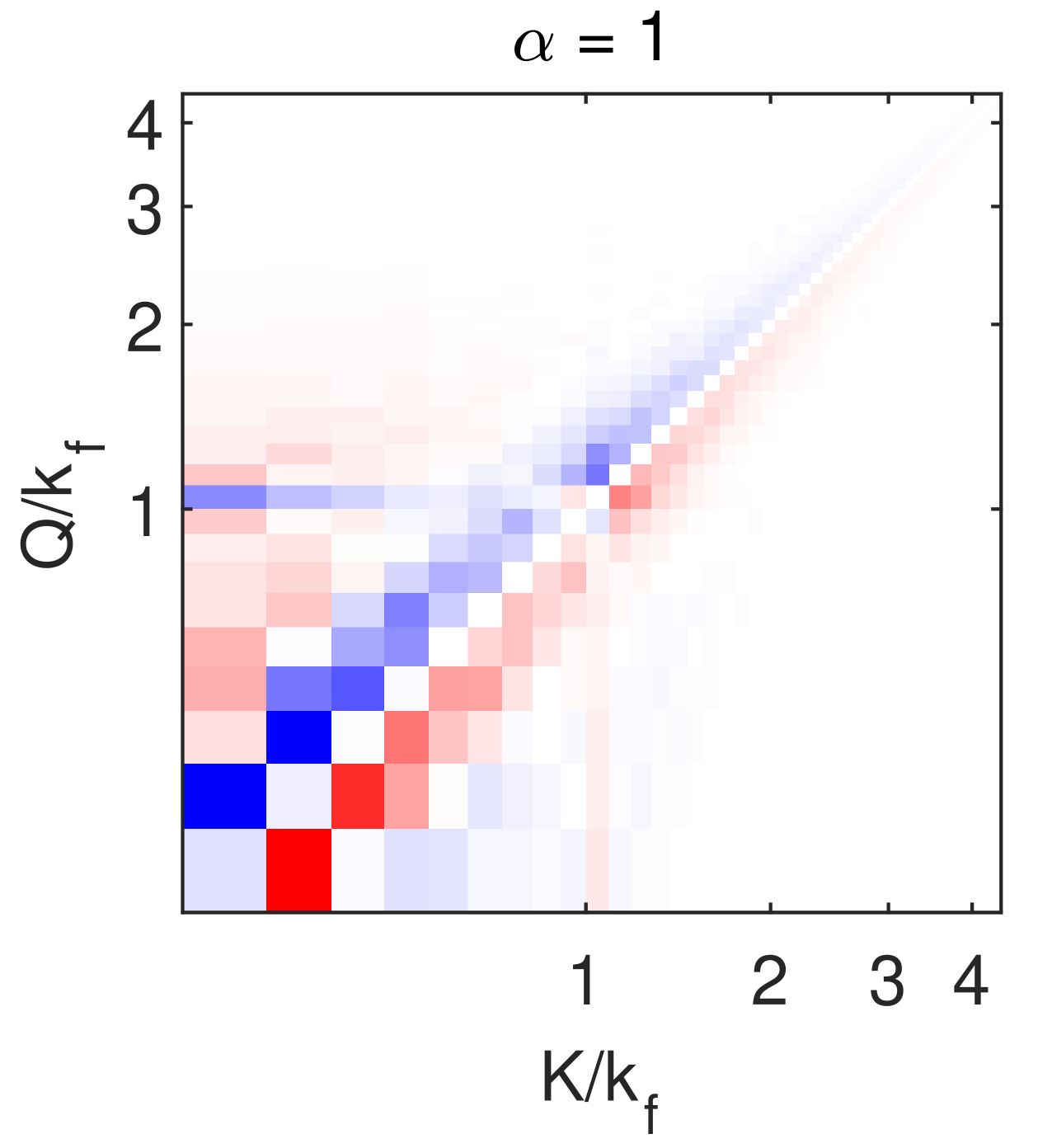}
		\caption{}
	\end{subfigure}
	\begin{subfigure}[h]{0.24\textwidth}
		\includegraphics[width=\linewidth]{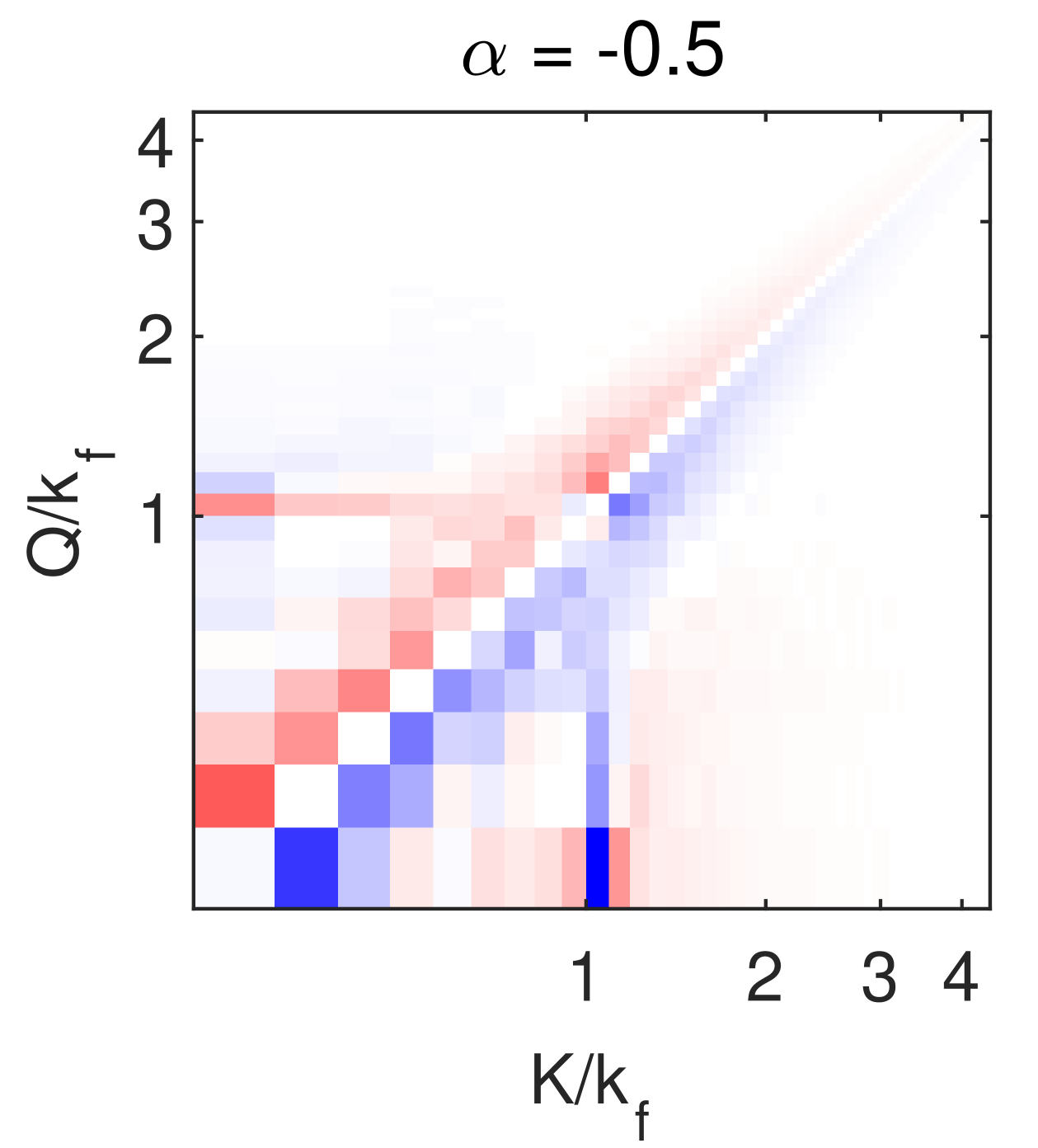}
		\caption{}
	\end{subfigure}
	\begin{subfigure}[h]{0.24\textwidth}
		\includegraphics[width=1.2\linewidth]{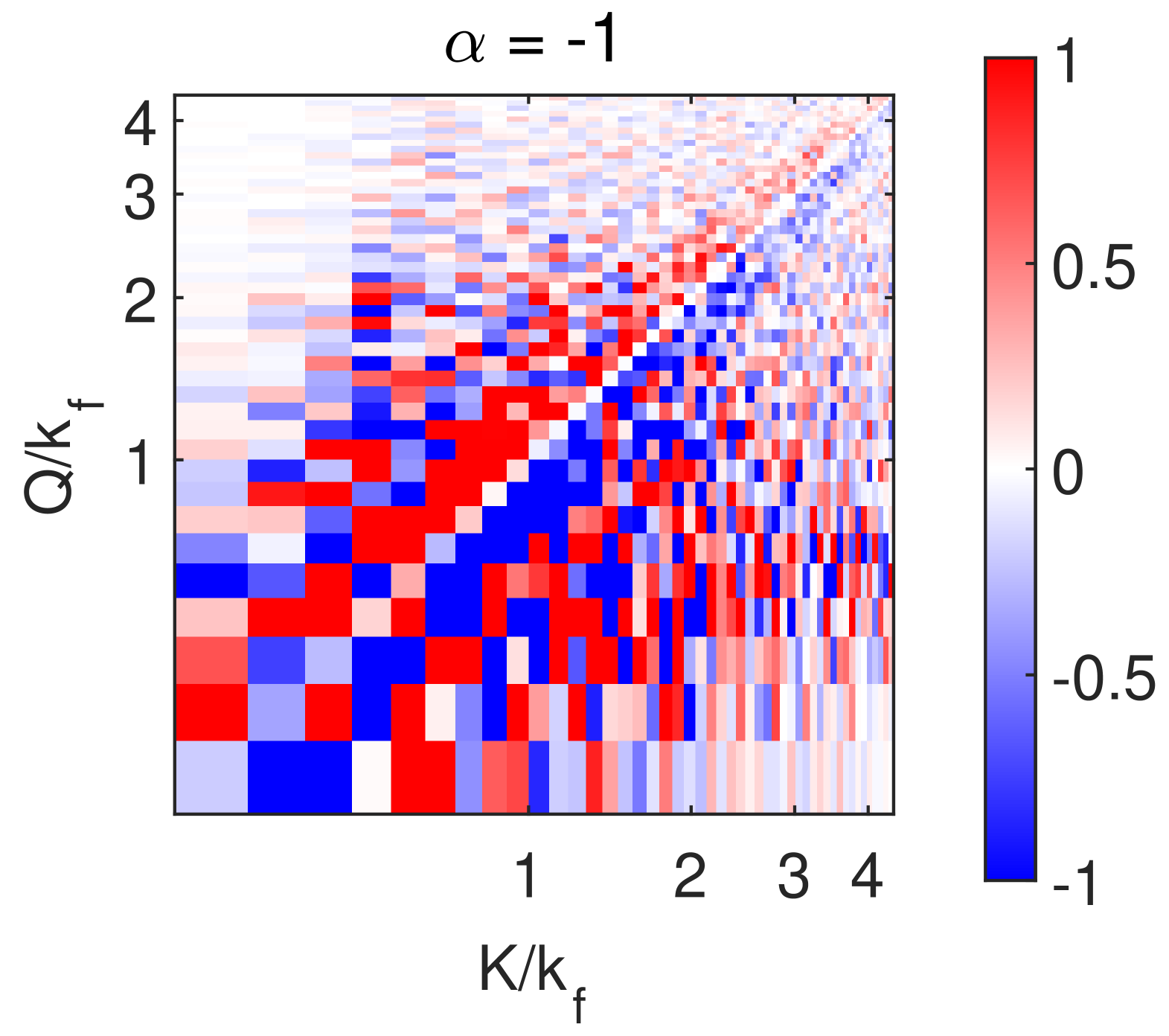}
		\caption{}
	\end{subfigure}
	%
	\begin{subfigure}[h]{0.24\textwidth}
		\includegraphics[width=\linewidth]{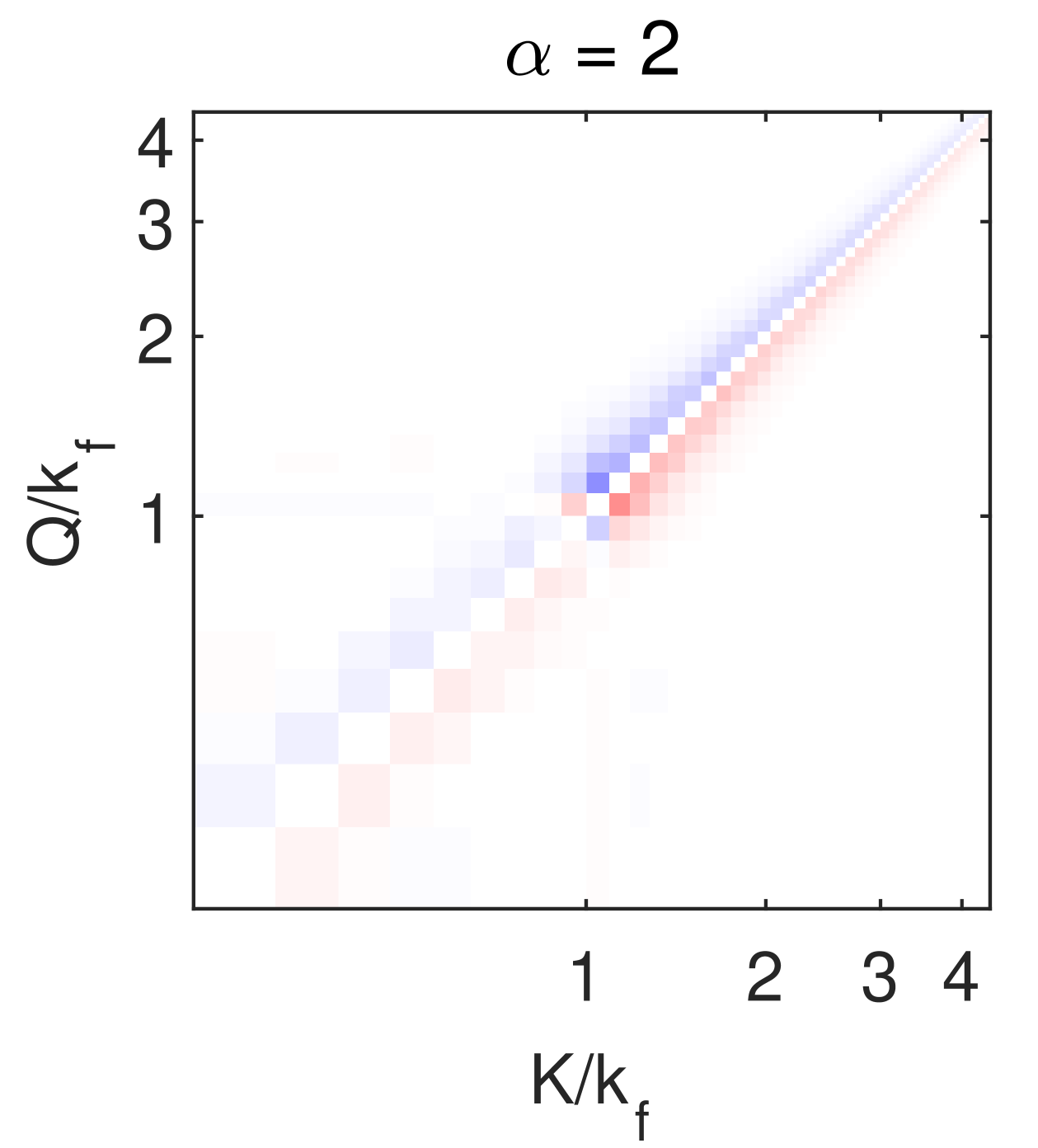}
		\caption{}
	\end{subfigure}
	\begin{subfigure}[h]{0.24\textwidth}
		\includegraphics[width=\linewidth]{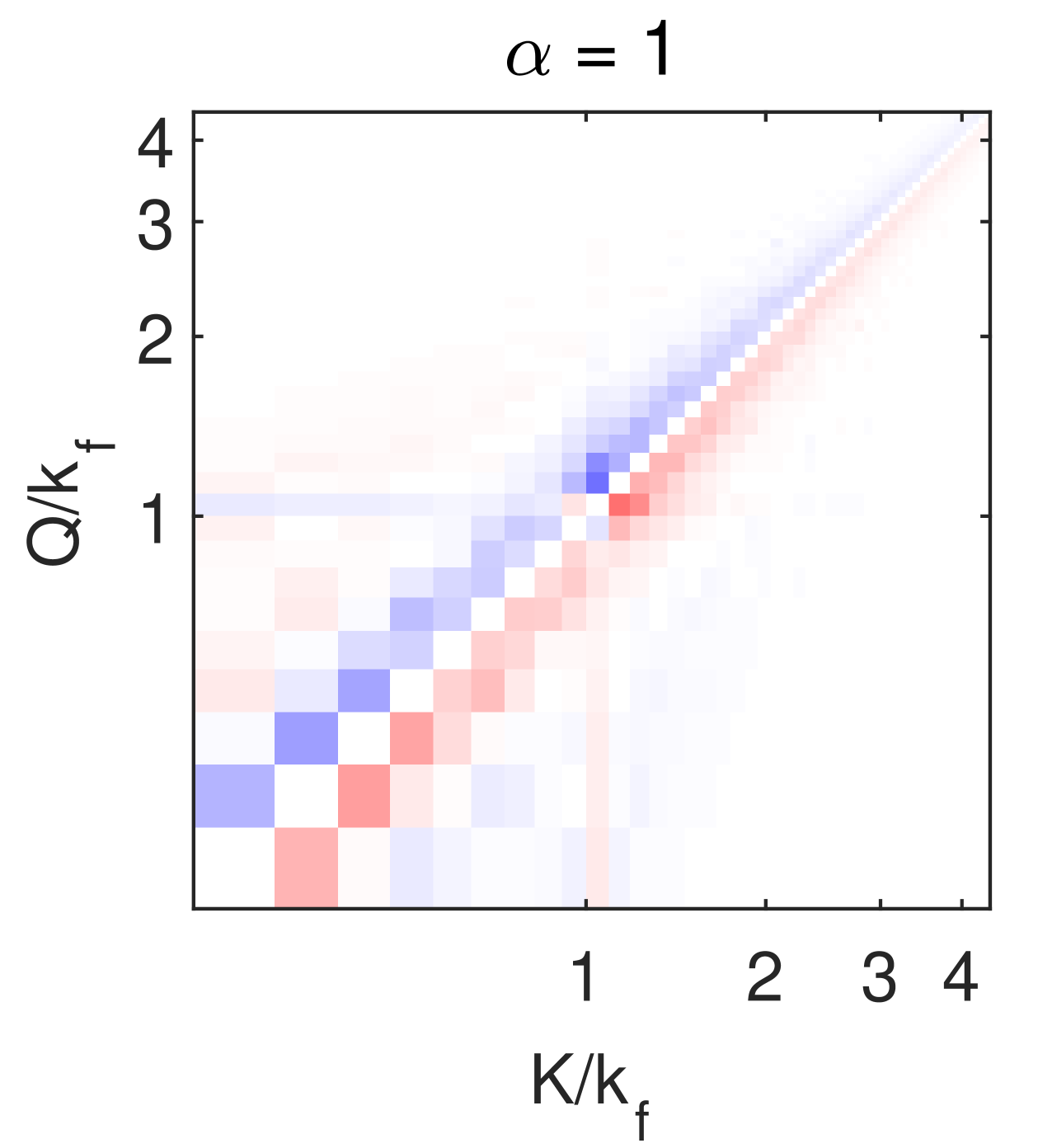}
		\caption{}
	\end{subfigure}
	\begin{subfigure}[h]{0.24\textwidth}
		\includegraphics[width=\linewidth]{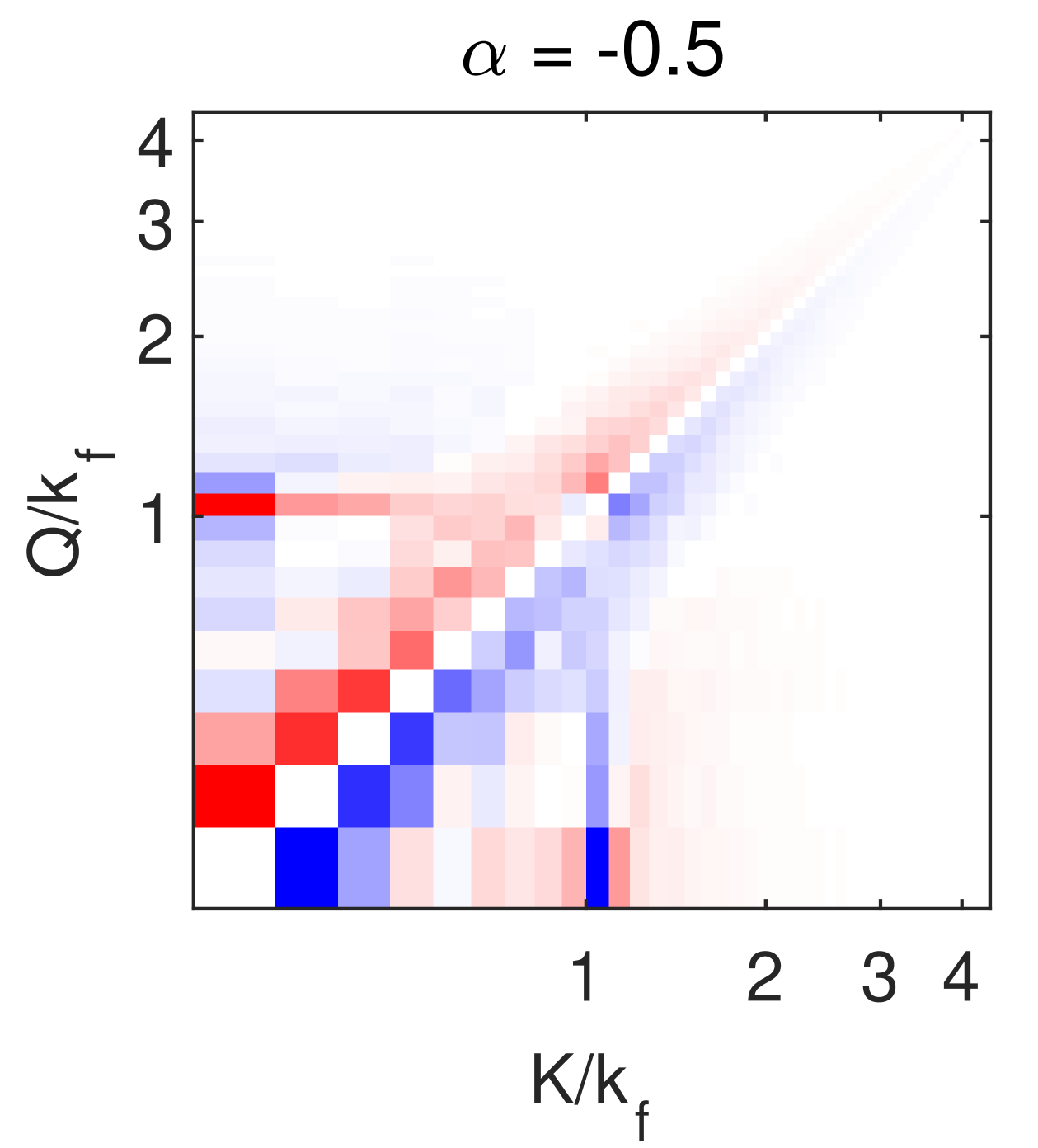}
		\caption{}
	\end{subfigure}
	\begin{subfigure}[h]{0.24\textwidth}
		\includegraphics[width=1.2\linewidth]{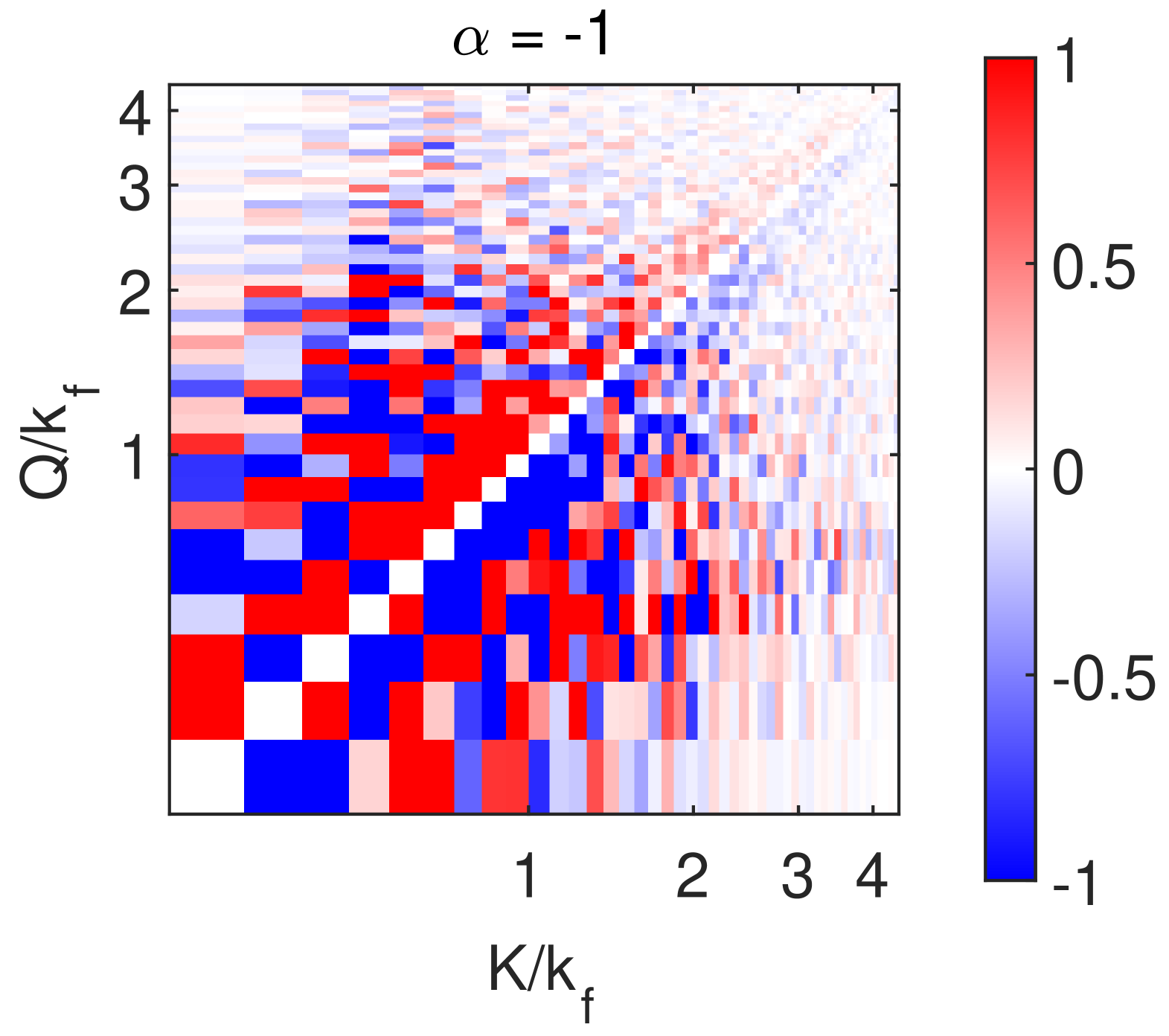}
		\caption{}
	\end{subfigure}
%
\caption{\label{fig:shell_to_shell_transfers} The top row shows shell-to-shell transfers of generalised energy as $T_E(K,Q)/\epsilon$ for 
		(a) $\alpha=2.0$, (b) $\alpha=1.0$, (c) $\alpha=-0.5$, and (d) $\alpha=-1.0$. 
		The bottom row shows shell-to-shell transfers of generalised enstrophy as $T_\Omega (K,Q)/\xi$ for 
		(e) $\alpha=2.0$, (f) $\alpha=1.0$, (g) $\alpha=-0.5$, and (h) $\alpha=-1.0$. 
		The colorbars are the same for all the plots and are shown only in plots d) and h) for clarity. The shell-to-shell transfers are obtained from DNS with $Re_+=1381$, $Re_-=10^6$ and $k_f=8\sqrt{2}$. \new{The shell-to-shell transfers are averaged \revi{over at least} 100 realisations \vd{after a stationary state is achieved} for all values of $\alpha$}.}	
\end{figure}

We quantify the locality of the nonlinear triadic interactions by analysing the transfer rate of generalised energy transferred via the generalised vorticity advection term from one shell of wavenumbers $Q < k < Q + \Delta k$ to another shell of wavenumbers $K < k < K + \Delta k$. We define this shell-to-shell transfer function as 
\begin{equation}
T_E(K,Q,t) = \avg{\psi_K ( \bold u \cdot \nabla q_Q )},
\label{eq:s2sE}
\end{equation}
where $\psi_K$ and $q_Q$ are the streamfunction and generalised vorticity fields filtered such that only the wavenumbers at shell $K$ and $Q$ are kept, respectively. The transfer term $T_E(K,Q,t)$ conserves the generalised energy, i.e. it does not generate or destroy $E_G$ but it is responsible for the redistribution of the generalised energy across different scales. \vbj{By construction, \vd{Eq. \eqref{eq:s2sE} transfers} energy from \vd{$\psi_K$ modes to $q_Q$ modes} and hence it is not anti-symmetric.  
This is also evident from the numerical results of the top row in Fig.\ref{fig:shell_to_shell_transfers}. 
On the other hand, \vdd{the shell-to-shell transfer of generalised enstrophy, from $q_K$ modes to $q_Q$ modes, is defined as}
\begin{equation}
	T_\Omega (K, Q, t) = \avg{q_K ( \bold u \cdot \nabla q_Q )}.
\end{equation}
$T_\Omega$ is anti-symmetric under exchange of $K$ and $Q$. This is also verified numerically (see bottom row of Fig. \ref{fig:shell_to_shell_transfers}).}

The contour plots in the top row of Fig. \ref{fig:shell_to_shell_transfers} show the  value of the time averaged shell-to-shell transfer function $T_E(K,Q)$ normalised by the energy injection rate $\epsilon$ for a) $\alpha = 2.0$ b) $\alpha = 1.0$, c) $\alpha = -0.5$ and d) $\alpha = -1.0$. \vbj{The $K/k_f$ and $Q/k_f$ axes are shown in logarithmic scale for better clarity \vdd{of the range $K/k_f <1$ and $Q/k_f < 1$}.}
\vbj{Local shell-to-shell transfers occur close to the diagonal $K=Q$, while non-local transfers occur away from the diagonal $K=Q$.} For $\alpha = 2$ the shell-to-shell transfer \vd{$T_E(K,Q)/\epsilon$} is mostly concentrated at $K/k_f < 1$, $Q/k_f < 1$ and close to the diagonal $K = Q$, indicating that generalised energy cascades to large scales and mostly via local \vbj{transfers}. For values of $\alpha < 2$, we observe that the transfer becomes more significant at $K/k_f > 1$ and $Q/k_f > 1$ and this is because the \vdd{cascade of $E_G$ transitions 
gradually} to small scales. Moreover, for $\alpha = 1$ and $-0.5$ the shell-to-shell transfer \vd{$T_E(K,Q)/\epsilon$} happens not only locally close to $K = Q$ but there are also two non-local branches, which become more and more significant as $\alpha \to -0.5$. These are the vertical branch, 
\vdd{close to $K/k_f = 1$ along the wavenumbers $Q/k_f \lesssim 1$
and the horizontal branch, 
close to $Q/k_f = 1$ along the wavenumbers $K/k_f \lesssim 1$.}
At $\alpha = -1$, we find that \vdd{$T_E(K,Q)/\epsilon$} is fully non-local spanning the whole range of wavenumbers. 

In a similar fashion the contour plots in the bottom row of Fig. \ref{fig:shell_to_shell_transfers} \vbj{show the time averaged} shell-to-shell transfer function $T_\Omega(K,Q)$ normalised by the \vbj{enstrophy} injection rate $\xi$ for e) $\alpha = 2.0$ f) $\alpha = 1.0$, g) $\alpha = -0.5$ and h) $\alpha = -1.0$.  
For $\alpha = 2$ the shell-to-shell transfer \vd{$T_\Omega(K,Q)/\xi$} occurs predominantly at $K/k_f > 1$, $Q/k_f > 1$ and close to the diagonal $K = Q$, indicating that generalised enstrophy cascades to small scales \revi{through} local interactions in wavenumber \vdd{space}. For values of $\alpha < 2$, we observe that \vd{$T_\Omega(K,Q)/\xi$} turns gradually to low wavenumbers, i.e. at $K/k_f < 1$ and $Q/k_f < 1$ and this happens since the cascade of \vd{$\Omega_G$} transitions to large scales. In addition, for $\alpha = 1$ and $-0.5$ the shell-to-shell transfer happens via three branches. One local branch \vd{adjacent to} $K = Q$ and two non-local branches, which become more and more significant as $\alpha \to -0.5$. Again these branches are the vertical  and horizontal branches that have similar pattern to the non-local branches of $T_E(K,Q)/\epsilon$. Finally, at $\alpha = -1$, we also find that \vd{$T_\Omega(K,Q)/\xi$} is fully non-local \vbj{spread} across the wavenumbers. 

The fact that the cascades of the generalised energy and enstrophy become gradually non-local as we go from positive to negative values of $\alpha$ is a notion clearly at variance with the typical locality assumption of the Kolmogorov phenomenology. This might be a reason why the energy spectra in Fig. \ref{fig:Spectra} do not agree with the scaling predictions from Eqs. \eqref{eqn:spec_pred1} and \eqref{eqn:spec_pred2}.

\section{Theoretical estimates}
\label{sec:bounds_discussion}

In this section we discuss theoretical bounds on generalised energy and enstrophy dissipation rates, which we derive in appendix \ref{sec:bounds} using exact mathematical inequalities. As we already described in section \ref{sec:results} the dissipation rates \revi{did not reach Reynolds-number-independent behaviour in our DNS, even at the highest Reynolds numbers examined (which required high spatial resolutions)}. Thus, we attempt to study the infinite Reynolds number behaviour of the dissipation rates theoretically, at \vdd{the} statistically stationary regime, using bounds. Moreover, we also consider the condensate regime, 
\vdd{where reaching statistical stationarity is computationally very challenging}, by ignoring the large scale dissipation term.

Note that numerically we used hyper and hypo-viscous dissipation terms to reach high enough Reynolds numbers. \revi{Theoretically, though, we are} not limited by computational power, hence we derive the bounds using the standard dissipation terms. So, let us start with Eq. \eqref{eq:qmodel} and consider the small and large scale dissipation terms with $n = 1$ and $m = 0$, respectively, given by
	\begin{equation}
		\partial_t q + \revi{J(q, \psi)} = \vdd{\nu_+} \nabla^2 q -\nu_- q + f_q,
		\label{eq:qmodel_stand}
	\end{equation}	
where $f_q$ is a time-independent external forcing and broadband in the spectral domain. \vdd{The above equation is studied in the domain $[0,2 \pi] \cross [0,2 \pi ]$ subject to periodic boundary conditions.} The corresponding \vd{evolution equation for the streamfunction} can be found using Eq. \eqref{eq:psiq}, i.e.
\begin{equation}
 \partial_t \psi + \revi{J'(q,\psi)} = \nu_+\nabla^2 \psi - \nu_- \psi +  f_\psi,
 \label{eq:bounds2_m}
\end{equation}
where \revi{$f_\psi = (-\nabla^2)^{-\alpha/2} f_q$, and $J'(q,\psi) = (-\nabla^2)^{-\alpha/2}J(q,\psi)$}. \revi{We define the forcing wavenumber as, } $k_f = (\avg{|\nabla^2 f_\psi|^2} / \avg{|f_\psi|^2})^{1/4}$, and
the dimensionless measures of the generalised energy and enstrophy dissipation rates as
	\begin{equation}
		c_\epsilon = \frac{\epsilon_+}{U^3k_f^{\alpha-1}}, \quad\quad 
		c_\xi = \frac{\xi_+}{U^3k_f^{2\alpha-1}},
		\label{eq:c_eps_c_xi_m}
	\end{equation}
	\vd{where $U = \avg{|\bold u|^2}^{1/2} = \avg{|{\nabla} \psi|^2}^{1/2}$ is the rms velocity and $k_f$ is used for dimensional consistency.}
For the purposes of the \vdd{analysis in appendix \ref{sec:bounds}}, it is more useful and convenient to define the Reynolds number based on the rms velocity instead of the forcing amplitude. This leads to the following definitions,
	\begin{equation}
		\Re = \frac{U}{k_f \nu_+}, \quad\quad 
		\Rh = \frac{Uk_f}{\nu_-}.
		\label{eq:Reynolds_m}
	\end{equation}
\vd{In what follows, we discuss the main results from the bounds. The technical details of the derivation are described in appendix \ref{sec:bounds}. The $\alpha > 0$ and $\alpha < 0$ cases are discussed separately owing to the nature of the cascades, which is governed by the sign of $\alpha$. Mathematically the two cases arise because for $\alpha > 0$ generalised enstrophy is a higher order derivative quantity than generalised energy and vice versa for $\alpha < 0$. The idea for the bounds is that once the \vdd{dissipative term with the highest order derivative is bounded, we then proceed to bound the other dissipative term} (see appendix \ref{sec:bounds}).}

\subsection{Positive $\alpha$}
	Using the generalised enstrophy balance and \nnew{assuming a statistically stationary state} we are able to bound the small scale dissipation rate of generalised enstrophy (see appendix \ref{sec:bounds}) as,
    \revi{
    \begin{align}
	\xi_+ \leq C_0k_f^{2\alpha-1} U (C_1^{'} U^2 + C_2^{'} \nu_+ k_f U + C_3^{'}\nu_-k_f^{-1}U),
	\label{eq:xi_plus}
	\end{align}
    }
where \revi{$C_0, C_1^{'}, C_2^{'}, C_3^{'}$} are constants that depend on the form of the forcing function and the domain geometry. The above equation in non-dimensional form can be written as,
\begin{align}
	c_\xi \leq C_1 + C_2 \Re^{-1} + C_3 \Rh^{-1},
	\label{eq:c_xi}
\end{align}
where we have divided by $k_f^{2\alpha - 1} U^3$ and used the definitions \eqref{eq:c_eps_c_xi_m} and \eqref{eq:Reynolds_m}. 
So, as \vdd{$\Re \to \infty$ and $\Rh \to \infty$}, we find that for \final{$0<\alpha\leq 1$ and $\alpha=2$} the generalised enstrophy dissipation rate $c_\xi$ is bounded from above by a positive constant, which is independent \revi{of} $\Re$ and $\Rh$. 
\vdd{This bound 
is an upper bound, 
which allows both for zero and finite $c_\xi$ as a possible solution. So, \eqref{eq:c_xi} does not rule out a dissipation anomaly for the generalised enstrophy dissipation rate in the infinite Reynolds number limit.}
 
Then, we proceed to bound the energy dissipation rate, where its dimensionless form is found to obey
\begin{align}
 c_{\epsilon} \leq \Re^{-1/2}(C_5 + C_6 \Re^{-1} + C_7 \Rh^{-1})^{1/2}.
  \label{eq:c_eps}
\end{align}
In this case, as \vdd{$\Re \to \infty$ we get $c_{\epsilon} \to 0$, since $c_\epsilon$ is positive definite}, implying no forward cascade of generalised energy for \final{$0<\alpha\leq1$ and $\alpha=2$} in the infinite Reynolds number limit.

If we now consider no large scale dissipation, a \nnew{large-scale condensate} is formed for $\alpha > 0$. In this case, \final{considering 
$\Rh\to \infty$ and  $\Re\to\infty$} in equations \eqref{eq:c_eps} and \eqref{eq:c_xi}, we can claim that $c_{\epsilon} \to 0$ \nnew{and the upper bound on $c_{\xi} $ can allow for cascade solutions to exist.}

\subsection{Negative $\alpha$}
Using the generalised energy balance \nnew{and assuming a statistically stationary state} we are able to bound the small scale dissipation rate of generalised enstrophy (see appendix \ref{sec:bounds}) as
\revi{
\begin{align}
	\epsilon_+ &\leq Ck_f^{\alpha -1} U (\tilde{C}_{1}^{'} U^2 + \tilde{C}_{2}^{'} \nu_+ k_f U + \tilde{C}_{3}^{'}\nu_-k_f^{-1}U),
	\label{bounds19}
\end{align}
}
where \revi{$C,\tilde{C}_1^{'}, \tilde{C}_2^{'}, \tilde{C}_3^{'}$} are constants which depend on the forcing function and the domain parameters. Expressing the above equation in non-dimensional form gives,
\begin{align}
	c_{\epsilon} &\leq \tilde{C}_{1} + \tilde{C}_{2} \Re^{-1} + \tilde{C}_{3} \Rh^{-1}
	\label{eq:c_eps_minus}
\end{align}
where we have divided by $k_f^{\alpha - 1} U^3$ and used the definitions \eqref{eq:c_eps_c_xi_m} and \eqref{eq:Reynolds_m}.
Thus, as \vdd{$\Re \to \infty$ and $\Rh \to \infty$} we find that for $\alpha < 0$, the generalised energy dissipation rate $c_{\epsilon}$ is bounded from above by a constant independent of $\Re$ and $\Rh$. \vdd{Again, this expression is an upper bound, which allows both for zero and finite $c_{\epsilon}$ as a possible solution. So, \eqref{eq:c_eps_minus} does not rule out a dissipation anomaly for the generalised energy dissipation rate in the infinite Reynolds number limit.}
	
Next we find that the non-dimensional generalised enstrophy dissipation rate obeys, 
\begin{align}
 c_{\xi} \leq \Re^{-1/2}(\tilde{C}_{5} + \tilde{C}_{6} \Re^{-1}  + \tilde{C}_{7}\Rh^{-1})^{1/2}.
 \label{eq:c_xi_minus}
\end{align}
Thus, \vdd{as $\Re \to \infty$ we get $c_\xi \to 0$, since $c_\xi$ is positive definite, implying} no forward cascade of generalised enstrophy for $\alpha < 0$ in the infinite Reynolds number limit.

\vdd{Again, if} we now consider no large scale dissipation, a \nnew{large-scale condensate can be} formed for $\alpha < 0$. \final{Taking the limits 
$\Rh\to\infty$, and $\Re\to\infty$} in equations \eqref{eq:c_eps_minus} and \eqref{eq:c_xi_minus}, we can claim that \nnew{$c_{\xi} \to 0$ and 
 the upper bound on $c_{\epsilon}$ can allow for cascade solutions to exist.}

\ks{In summary, the bounds we derived for positive and negative $\alpha$ demonstrate that there is a transition of cascades between the generalised energy and enstrophy, if the upper bounds are taken to be exact. In other words, for $\alpha > 0$ the bounds suggest that in the infinite Reynolds number \nnew{limit} $\epsilon_+ \to 0$ possibly due to the inverse cascade, and $\xi_+$ can remain finite due to the dissipation anomaly of the forward cascade, while the opposite is true for $\alpha < 0$.}

\section{Conclusion}
\label{sec:conclusion}

In this paper we focus on the numerical analysis of the cascades in generalised 2D turbulence by systematically varying the parameter $\alpha$ of the model \vd{in the range $-1 \leq \alpha \leq 2$}. The value and sign of $\alpha$ \revi{determines whether the cascade of a given quadratic invariant is inverse, forward or bidirectional. We find that there is 
a transition in the direction of the cascades as one \vdd{varies} the value of $\alpha$}. At the threshold $\alpha = 0$ the nonlinear term vanishes and the flow is laminar. We find that as we move away from $\alpha = 0$ the laminar flow undergoes a linear small wavelength instability at a critical value $\alpha_c$, which is the same both for positive and negative $\alpha$. We observe that $\alpha_c \approx const.$ as we increase $Re_-$, however \vd{$\alpha_c \propto Re_+^{-2}$} as $Re_+$ increases. This scaling is also verified theoretically by doing a perturbation analysis. 

Using mathematical inequalities for \final{$0<\alpha \leq 1$ and $\alpha=2$} we are able to bound the dimensionless dissipation rates of generalised energy \vbj{$c_{\epsilon} \equiv \epsilon/(U^3 k_f^{\alpha -1}) \leq \Re^{-1/2}(C_5 + C_6 \Re^{-1} + C_7 \Rh^{-1})^{1/2}$} and enstrophy \vbj{$c_{\xi} \equiv \xi/ (U^3 k_f^{2\alpha - 1}) \leq C_1 + C_2 \Re^{-1} + C_3 \Rh^{-1}$}. 
\vdd{In the limit of $\Re \to \infty$, these bounds state that $c_{\epsilon} \to 0$,} 
\revi{which can be related to the inverse cascade of $E_G$}, while $c_{\xi}$ can either become \vdd{zero or finite 
due to the dissipation anomaly of the forward cascade of $\Omega_G$}. 
On the other hand, for $\alpha < 0$ we obtain the bounds \vbj{$c_{\epsilon} \leq \tilde{C}_1 + \tilde{C}_2 \Re^{-1} + \tilde{C}_3 \Rh^{-1}$} and \vbj{$c_{\xi} \leq \Re^{-1/2}(\tilde{C}_5 + \tilde{C}_{6} \Re^{-1} + \tilde{C}_{7} \Rh^{-1})^{1/2}$}. 
\vdd{In this case for $\Re \to \infty$, the bounds give $c_{\xi} \to 0$, \revi{which can be related to the inverse cascade of $\Omega_G$}, while $c_{\epsilon}$ can either become zero or finite 
\vdd{due to} the dissipation anomaly of the forward cascade of $E_G$.}

In numerical simulations
when $\alpha > 0$ the generalised energy $E_G$ cascades inversely while generalised enstrophy $\Omega_G$ cascades forward and the opposite happens when $\alpha < 0$ (see Fig. \ref{fig:transitions}). Moreover, we find that the amount of dissipation rates of $E_G$ and $\Omega_G$ depend on the system parameters $Re_-$ and $Re_+$. This transition from positive to negative \vd{$\alpha$} is also clear from the spectral fluxes (see Fig. \ref{fig:fluxes}), 
\vd{which swap signs, indicating the transition of the cascades' direction.}
The energy spectra at large scales, i.e. at length scales larger than the forcing length scale, seem to agree with Kolmogorov type scalings of the generalised 2D turbulence. However, this is not true for length scales smaller than the forcing length scale. \nnew{A reason for this} discrepancy might be that higher resolution computations are required to shed light or that the nonlocal transfers of generalised energy and enstrophy might be the cause of this disagreement. Finally, it is interesting to mention that due to the forward cascade of $E_G$ for $\alpha < 0$, the classical vorticity field, which is dominated by vortex filaments, is reminiscent of the filaments observed in 3D Navier-Stokes turbulence. 

This study shows that the generalised model of 2D turbulence is another \vdd{convenient} set-up to study the transition of turbulent cascades. \revi{However, this model is unique as 
\vdd{it} shows features} that appear in 3D turbulence, like the vortex filaments, emerging in two dimensions. \vd{So, further numerical studies at higher Reynolds numbers are required to deepen our understanding and relate the dynamics of the vortex filaments in two dimensions to the dissipation anomaly of generalised energy, if present, for negative values of $\alpha$.} Further mathematical analysis of this model dissipation anomaly for $\alpha < 0$ might provide new insights \final{into} the regularity of solutions of the 3D Navier-Stokes equations. The study of locality of cascades using the shell-to-shell transfers from observational and experimental data of geophysical relevance would complement our findings, e.g. for the SQG model ($\alpha = 1$), and the non-locality indicates that we should be thinking beyond the Kolmogorov type of arguments for these models.

\section{Acknowledgment}
We wish to thank Krishna Kumar for useful discussions and support. We also wish to thank the computing resources and support provided by PARAM Shakti supercomputing facility of IIT Kharagpur established under National Supercomputing Mission (NSM), Government of India and supported by Centre for Development of Advanced Computing (CDAC), Pune. We acknowledge support from NSM Grant No. DST/NSM/R\&D HPC Applications/2021/03.11, from the Institute Scheme from Innovative Research and Development (ISIRD), IIT Kharagpur, Grant No. IIT/SRIC/ISIRD/2021–2022/08 and the Start-up Research Grant No. SRG/2021/001229 from Science \& Engineering Research Board (SERB), India.

\section{Declaration of Interests}
The authors report no conflict of interest.

\appendix
\section{Numerical Setup} \label{App:num_setup}

We perform \vdd{DNS} in a periodic square domain 
by numerically integrating 
\vd{the generalised vorticity equations \eqref{eq:qmodel}-\eqref{eq:psiq}}
using the pseudospectral method \citep{GottliebOrszag1977}. We decompose the streamfunction $\psi(x,y,t)$ into basis functions of Fourier modes, viz.
\begin{align}
	\label{eq:Basis}
	\psi({\bf x},t) &= \sum^{N/2}_{{\bf k} = -N/2} \widehat{ \psi}_{\bf k}(t) e^{i {\bf k} \cdot \mathbf{x}},
\end{align}
where $\widehat{ \psi}_{\bf k}$ is the amplitude of the ${\bf k} = (k_x, k_y)$ mode of $\psi$, and $N$ denotes the number of aliased modes in the $x$ and $y$ directions. A third-order Runge-Kutta scheme is used for time advancement. 
The aliasing errors are removed with the 2/3 rule, which implies that the maximum wavenumber $k_{max} = N/3$. 
\vd{
The numerical parameters we considered are listed in Table \ref{tbl:parameters}.
Time-averaged quantities are computed once the system has reached a statistically stationary regime. 
The computations were performed on graphics cards (GPUs), which provided three times speedup in contrast to computations on processors (CPUs). 
}

\begin{table}
	\makebox[\textwidth][c]{
		\begin{tabular}{cccccccc}
			&       & \multicolumn{2}{c}{a)$Re_+=1381$} & \multicolumn{4}{c}{b)$Re_-=4\times10^7$} \\
			\cmidrule(lr){2-4}\cmidrule(lr){5-8}
			& $Re_-^{(1)}=10^6$  & $Re_-^{(2)}=4\times10^7$  & $Re_-^{(3)}=2\times10^9$  & $Re_+^{(1)}=345$  & $Re_+^{(2)}=1381$ & $Re_+^{(3)}=3450$ & $Re_+^{(4)}=6900$\\ \midrule
			N & $512^2$ & $1024^2$ & $2048^2$ & $1024^2$ & $1024^2$ & $1024^2$ & $1024^2$ \\
			$k_f$ & $8\sqrt{2}$ & $16\sqrt{2}$ & $32\sqrt{2}$ & $16\sqrt{2}$ & $16\sqrt{2}$ & $16\sqrt{2}$ & $16\sqrt{2}$ \\
		\end{tabular}%
	}
	\caption{Numerical parameters for the two set of runs corresponding to a) fixed $Re_+$ with varying $Re_-$ and b) fixed $Re_-$ with varying $Re_+$. \final{$N$ is the number of Fourier modes in our computations}, $Re_+$ and $Re_-$ are small and large scale Reynolds numbers and $k_f$ is the forcing wavenumber.}
	\label{tbl:parameters}%
\end{table}%

\section{Perturbation Analysis}
\label{sec:scaling_law}

In this section, we look at the solutions 
\vdd{from $\alpha = 0$ up to the critical point $\alpha_c$.} \revi{This region} corresponds to the no-flux solution since the nonlinearity in the governing equation vanishes. As we go away from $\alpha_c$, we can do a perturbation analysis knowing that the nonlinear terms are small compared to the linear dissipation terms, with $\alpha$ being the small parameter. \revi{For $|\alpha| \ll 1$}, the solution is given by a laminar base flow $\psi_b$. This is the base flow around which the perturbation analysis will be carried out to find the 
\vdd{$\alpha_c$} at which this solution loses stability. The laminar solution is obtained by balancing the hyper and hypo-viscous terms with the forcing, 
\begin{align}
\psi_b = \frac{f_0}{\nu_+ k_f^4 + \nu_{-} k_f^{-4}} \sin(k_f x) \sin(k_f y), \label{eq:base_flow}
\end{align}
which is the solution 
\revi{for $|\alpha| < \alpha_c$}.

As we 
\vdd{cross} the threshold $\alpha_c$, we write the fields as a sum of the base flow and a perturbation, $q = q_b + \widetilde{q}$, $\psi = \psi_b + \widetilde{\psi}$, where the terms with the subscript $(\cdot)_b$ denote the base flow and the terms with $\widetilde{(.)}$ denote the perturbation fields. 
Substituting this decomposition into the nonlinear term of equation \eqref{eq:qmodel} we get, 
\begin{align}
\revi{J (q,\psi) = J(q_b,\psi_b) + J(q_b,\widetilde{\psi}) + J (\widetilde{q}, \psi_b ) + J ( \widetilde{q}, \widetilde{\psi},).}
\end{align} 
The term \revi{$J ( q_b, \psi_b )$} is zero for the laminar solution since $q_b$ is linearly related to $\psi_b$ and for the linear stability analysis, the term \revi{$J ( \widetilde{q}, \widetilde{\psi} )$} can be neglected as the term is second order in the amplitude of perturbations. Thus, by retaining only the linear terms in the perturbation fields, the nonlinear term in the governing equation \eqref{eq:qmodel} becomes,
\begin{equation}
	\revi{J ( q, \psi ) \approx J ( q_b, \widetilde{\psi} ) + J ( \widetilde{q}, \psi_b ). }
	\label{eqn:nonlin_Jacob}
\end{equation}

\vd{
We first express the generalised vorticities $q_b$ and $\widetilde{q}$ in terms of their respective stream functions $\psi_b$ and $\widetilde{\psi}$. The forcing is at a single wavenumber, thus based on Eq. \eqref{eq:psiq} we can write 
\revi{
\begin{equation}
 \widehat{q_b}({\bf k}, t) = k_f^\alpha \widehat{\psi_b}({\bf k}, t),
 \label{eq:qb}
\end{equation}}
where the $\widehat{(\cdot)}$ denotes Fourier coefficients.
Similarly, the perturbation field $\widetilde{q}$ can be written as 
\begin{align}
\widehat{\widetilde{q}} ({\bf k}, t) &= k^\alpha \widehat{\widetilde{\psi}} ( {\bf k}, t) \nonumber \\ 
						 &= k_f^\alpha (1 + \alpha \log( \frac{k}{k_f} ) + \mathcal{O}(\alpha^2) ) \widehat{\widetilde{\psi}}( {\bf k}, t) \nonumber \\
						 &= k_f^\alpha \widehat{\widetilde{\psi}}( {\bf k}, t) + \alpha \widehat{\widetilde{Q}}( {\bf k}, t) + \mathcal{O}(\alpha^2)
\end{align}
with $k^\alpha = k_f^\alpha (1 + \alpha \log( \frac{k}{k_f} ) + O(\alpha^2))$ expanded in the limit of $\alpha \rightarrow 0$ and 
\begin{align}
\widehat{\widetilde{Q}} ({\bf k}, t) = k_f^\alpha \log (\frac{k}{k_f}) \widehat{\widetilde{\psi}}\vdd{({\bf k}, t)}. \label{eq:expansion_phi}
\end{align}
Using Eq. \eqref{eq:qb} and Eq. \eqref{eq:expansion_phi} in real space and substituting them into \eqref{eqn:nonlin_Jacob} we get,
\begin{align}
\revi{J ( q, \psi ) \approx k_f^\alpha J ( \psi_b, \widetilde{\psi} ) + k_f^\alpha J ( \widetilde{\psi}, \psi_b ) + \alpha J (\widetilde{Q}, \psi_b)}
\end{align}
The first two terms cancel out each other with only the third term remaining which is proportional to $\alpha$. 
}

Now we can derive a scaling \vdd{for} $\alpha_c$ with $Re_+$ from the balance between the non-linear term $\alpha \revi{J(\widetilde{Q}, \psi_b)}$ and the hyper-viscous dissipation term $-\nu_+ (-\nabla^2)^{n} \widetilde{\psi}$ with $n = 2$. At the threshold $\alpha_c$ of the linear instability we use Eqs. \eqref{eq:base_flow} and \eqref{eq:expansion_phi} to find the following leading order scaling
\begin{align}
	\revi{J(q,\psi)} &\propto \alpha_c \frac{f_0}{\nu_+ k_f^2} \; g_1\left(\frac{k_i}{k_f}\right) \widetilde{\psi}, 
	\label{LHS} \\
	\nu_+ \nabla^{4} \widetilde{\psi} &\propto \nu_+ k_f^4 \; g_2\left(\frac{k_i}{k_f}\right) \widetilde{\psi},
	\label{RHS}
\end{align}
where $k_i$ denotes the typical wavenumber of the unstable mode $\widetilde{\psi}$ and $g_1, g_2$ are two functions of $k_i/k_f$. Equating \eqref{LHS} with \eqref{RHS} we get,
\begin{equation}
	\alpha_c \frac{f_0}{\nu_{+} k_f^2} \propto \nu_{+} k_f^4 \; g_3\left(\frac{k_i}{k_f}\right),
\end{equation}
where $g_3$ is a function of $k_i/k_f$. We find that the ratio $k_i/k_f$ remains almost constant as we vary $\Re_+$ over a few decades, which leads to the following scaling law for the threshold of the instability 
\begin{equation}
	\alpha_c \propto Re_{+}^{-2}.
\end{equation}

\section{Bounds} \label{sec:bounds}
Here we derive the upper bounds for the non-dimensionalised dissipation rates \vdd{defined} in Eq. \eqref{eq:c_eps_c_xi_m}. We start by reproducing the definitions from section \ref{sec:bounds_discussion}, with the Eq. \eqref{eq:qmodel_stand} reproduced below,
	\begin{equation}
		\partial_t q + \revi{J(q, \psi)} = \nu_+ \nabla^2 q  - \nu_- q + f_q,
		\label{eq:bounds1}
	\end{equation}
where \vd{$f_q$ is the external forcing function which is taken to be a general smooth function that is time independent}. \vd{The corresponding evolution equation for the streamfunction using Eq. \eqref{eq:psiq} is also reproduced below}, i.e.
\begin{equation}
 \partial_t \psi + \revi{J'(q, \psi)} = \nu_+\nabla^2 \psi - \nu_- \psi +  f_\psi,
 \label{eq:bounds2}
\end{equation}
where $f_\psi = (-\nabla^2)^{-\alpha/2} f_q$ and 
\vdd{the forcing wavenumber is defined as $k_f = (\avg{|\nabla^2 f_\psi|^2} / \avg{|f_\psi|^2})^{1/4}$.} We recall that, 
\revi{
\begin{equation}
 J'(q, \psi) = (-\nabla^2)^{-\alpha/2}J(q, \psi).
 \label{eq:Jprime}
 \end{equation}}
The dimensionless measures of the generalised energy and enstrophy dissipation rates are given by $c_{\epsilon}= \epsilon_+/(U^3k_f^{\alpha-1}), c_{\xi} = \xi_+/(U^3k_f^{2\alpha-1})$, taken from Eqs. \eqref{eq:c_eps_c_xi_m}.
	\vd{Here $U = \avg{|\bold u|^2}^{1/2} = \avg{|{\nabla} \psi|^2}^{1/2}$ is the rms velocity and $k_f$ is used for dimensional consistency.}
	The Reynolds number based on the rms velocity defined in Eqs. \eqref{eq:Reynolds_m} are given by, \new{$\Re = \frac{U}{k_f \nu_+}, \Rh = \frac{Uk_f}{\nu_-}$}.   Note that the Reynolds number defined here is based on \vdd{the} rms velocity.
To obtain the upper bound estimates, we start with bounding 
\vdd{
the dissipative term with the highest order derivative. From this estimate, we then bound the other dissipative term. Since the nature of the cascades is governed by the sign of $\alpha$, we consider the two cases separately.}

\subsection{Positive $\alpha$}
In the statistically stationary state, we have 
the generalised enstrophy balance\revi{,} $\xi_+ + \xi_- = \xi$, \revi{discussed in Eq. \eqref{eq:enstrophybal}, which can be \vdd{similarly}} derived from \eqref{eq:bounds1}, \vdd{where} $\xi_+ = \nu_+\avg{|\nabla q|^2}$, $\xi_- = \nu_- \avg{q^2}$ and $\xi = \avg{q f_q}$. From the enstrophy balance and Eq. \eqref{eq:coupling} we can write
\begin{align} 
 \xi_+ = \avg{q f_q} - \nu_- \avg{q^2} &\leq C^{'}k_f^{2\alpha} \avg{|\psi|^2}^{1/2} \avg{|f_{\psi}|^2}^{1/2} -\nu_-\avg{q^2} \\
 &\leq C^{'}k_f^{2\alpha}\langle|\psi|^2\rangle^{1/2}F_\psi,
 \label{eqn:zeta_bound1}
\end{align}
\vdd{where $F_{\psi} = \avg{|f_{\psi}|^2}^{1/2}$ is the forcing amplitude and $C^{'}$ is a constant.
Here}, we have applied the Cauchy-Schwarz inequality to the first term, $k_f^{2\alpha}\langle\psi f_{\psi}\rangle$,
%
and $\nu_-\langle q^2\rangle$ is a positive definite \vdd{quantity, hence it} can be neglected from the upper bound estimate. 
\vbj{Since $\psi$ is a lower derivative than the velocity field ${\bf u}$, we can use the \vd{Poincar\'e inequality \citep{doering1995applied}} to write}
\begin{align}
  \avg{|\psi|^2}^{1/2} &\leq C k_f^{-1}\avg{ |\bold u|^2}^{1/2} \nonumber \\
  &\leq C k_f^{-1} U,
  \label{eq:psi_rms}
\end{align} 
where $C$ is a constant that depends on the domain geometry. \vd{Combining Eq. \eqref{eqn:zeta_bound1} with Eq. \eqref{eq:psi_rms} yields a bound on the small scale dissipation rate of generalised enstrophy in terms of the forcing \vdd{amplitude}, i.e.
\revi{
\begin{equation}
    \xi_+ \leq C_0k_f^{2\alpha-1}UF_{\psi}.
    \label{eq:xi_bound}
\end{equation}
}
}

\vdd{where, $C_0$ is a constant. Now,} we would like to replace $F_\psi$ in terms of the \vdd{rms velocity,} which will help us obtain the bound \vdd{in terms of} the dimensionless dissipation rate $c_\xi$, \vdd{as it is defined based on the rms velocity}. 
For this, we follow \cite{alexakis2006energy} and multiply Eq. \eqref{eq:bounds2} by a smoothly varying, time independent function $\phi$, giving
\begin{align}
 \phi \partial_t \psi + \phi \revi{J'(q, \psi)} &= \nu_+\phi\nabla^2\psi - \nu_-\phi\psi + \phi f_\psi.
 \label{bounds4}
\end{align}
Performing spatiotemporal averaging and changing the order of differentiation using integration by parts, we get
\begin{equation}
\revi{ \avg{\phi J'(q, \psi)} } - \nu_+\avg{\psi \nabla^2\phi} + \nu_- \avg{\phi \psi} = \avg{\phi f_\psi},
 \label{bounds5}
\end{equation}
since $\avg{\phi \partial_t \psi} = 0$ \revi{due to statistical stationarity}.
Let us derive estimates for each term separately. For the second and third term on the left hand side of Eq. \eqref{bounds5}, we can use the Cauchy-Schwarz inequality and Eq. \eqref{eq:psi_rms} to write
\revi{
\begin{align}
 -\nu_+ \avg{\psi\nabla^2\phi} &\leq \nu_+ \avg{|\psi|^2}^{1/2} \avg{|\nabla^2\phi|^2}^{1/2} \nonumber \\
 					    &\leq \nu_+ C k_f^{-1} U \avg{|\nabla^2\phi|^2}^{1/2}, 
 \label{bounds7}
\end{align} }
and 
\begin{align}
 \nu_-\avg{\phi\psi} &\leq \nu_- \avg{|\phi|^2}^{1/2} \avg{|\psi|^2}^{1/2} \nonumber \\
                &\leq \nu_-\avg{|\phi|^2}^{1/2} C k_f^{-1} U. 
 \label{bounds7a}
\end{align}
For the first term of Eq. \eqref{bounds5}, using Eq. \eqref{eq:Jprime} we can write,
\begin{align}
\avg{\phi \revi{J'(q,\psi)}} &= \avg{\phi (-\nabla^2)^{-\frac{\alpha}{2}}\revi{J(q, \psi)}} \nonumber \\ 
			     &= \avg{\phi (-\nabla^2)^{-\frac{\alpha}{2}} (\bold{u} \cdot \nabla) q}.
        \label{eq:termJ1}
\end{align}
We proceed to simplify this expression by using the incompressibility condition of the velocity field and the periodicity of the domain. After some simplifications, we get
\final{
\begin{align}
  \avg{\phi J'(q,\psi)} &= \avg{\phi \nabla \cdot \left( (-\nabla^2)^{-\frac{\alpha}{2}}({\bf u}q) \right)} \nonumber \\
  &=-\avg{[(-\nabla^2)^{\frac{-\alpha}{2}}({\bf \nabla}\phi)].({\bf u}q)}
\end{align}
Now we apply H\"older's inequality to get, 
\begin{align}
\avg{\phi J'(q,\psi)} &\leq ||(-\nabla^2)^{-\alpha/2} \nabla\phi||_{\infty}||{\bf u}||_2||q||_2 \nonumber \\
&\leq ||(-\nabla^2)^{-\alpha/2} \nabla\phi||_{\infty}CU^2k_f^{-1+\alpha}
 \label{eq:psi_J2}
\end{align}}
\final{ where the $||.||_2$ term denotes the $L_2$ norm and 
$C$ is a constant that depends on the form of the forcing function and domain geometry. In the last step, we have used the relation $||q||_2\leq CUk_f^{-1+\alpha}$ for $0\leq\alpha\leq 1$. For $\alpha=2$, we obtain a similar bound as in \cite{alexakis2006energy}.}
The infinity norm is defined as $||\theta||_\infty = \max|\theta(x,y)|$, which gives the maximum absolute value of a function $\theta$ over the domain.

Putting Eq.\eqref{eq:psi_J2}, Eq.\eqref{bounds7} and Eq.\eqref{bounds7a} in Eq.\eqref{bounds5}, we get 
\begin{equation}
 \avg{\phi f_\psi} \leq 
 \final{ ||(-\nabla^2)^{-\frac{\alpha}{2}} \nabla \phi||_\infty Ck_f^{-1+\alpha} U^2}
			+ C \nu_+ \avg{|\nabla^2\phi|^2}^{1/2} k_f^{-1} U 
			+ C \nu_- \avg{|\phi|^2}^{1/2} k_f^{-1} U.
 \label{eq:phi_fpsi}
\end{equation}
By letting $\phi = f_\psi / F_\psi$ and using the definition $k_f^2 = \avg{|\nabla^2 f_\psi|^2}^{1/2} / \avg{|f_\psi|^2}^{1/2}$ we can simplify 
Eq. \eqref{eq:phi_fpsi} and obtain a bound relating the forcing amplitude $F_\psi$ and the \vdd{rms velocity}. \vdd{In the first term on the right hand side, the \final{infinity norm simplifies to $k_f^{1-\alpha}$ times a constant} which depends on the shape of the forcing function, the} second term in \vdd{the right hand side} becomes $C\nu_+k_f^2k_f^{-1}U$ and the last term becomes $C\nu_-k_f^{-1}U$, where the constants depend on the form of forcing function and geometry. 
\vdd{Eventually, Eq. \eqref{eq:phi_fpsi} turns into the following expression,}
\revi{
\begin{equation}
 F_\psi \leq C_1^{'} U^2 + C_2^{'} \nu_+ k_f U + C_3^{'}\nu_-k_f^{-1}U,
 \label{bounds9}
\end{equation}
}
where \revi{$C_1^{'}, C_2^{'}, C_3^{'}$} are constants that depend on the form of the forcing function and the domain geometry. 
Substituting \eqref{bounds9} in the expression for the generalised enstrophy dissipation rate \eqref{eq:xi_bound} we obtain the following bound,
\revi{
\begin{align}
	\xi_+ &\leq C_0k_f^{2\alpha-1} U (C_1^{'} U^2 + C_2^{'} \nu_+ k_f U \red{ + C_3^{'}\nu_-k_f^{-1}U}),
	\label{bounds10}
\end{align}
}
which in non-dimensional form can be written as,
\begin{align}
	c_\xi \leq C_1 + C_2\Re^{-1} \red{+ C_3\Rh^{-1}},
	\label{bounds11}
\end{align}
where we have divided by $k_f^{2\alpha - 1} U^3$ and used the definitions \eqref{eq:c_eps_c_xi_m} and \eqref{eq:Reynolds_m}.

\vdd{Now} we have obtained a bound on the \final{generalised} enstrophy dissipation rate, we then look to bound the generalised energy dissipation rate given by \revi{$\epsilon_+ = -\nu_+\langle\psi\nabla^2q\rangle$}. Using the Cauchy-Schwarz inequality, we get
\begin{align}
 \epsilon_+ &\leq \nu_+ \avg{|\nabla \psi|^2}^{1/2} \avg{|\nabla q|^2}^{1/2} \nonumber \\
		 &\leq \vbj{C_4} \nu_+^{1/2} U \xi_+^{1/2},
\end{align}
\vd{since $U = \avg{|\bold u|^2}^{1/2} = \avg{|{\nabla} \psi|^2}^{1/2}$ and $\xi_+ = \nu_+\avg{|\nabla q|^2}$.}
Using the bound for $\xi_+$ from Eq. \eqref{bounds10} \vbj{and dividing by $U^3k_f^{\alpha-1}$}, this inequality simplifies to
\begin{align}
	c_\epsilon \leq \Re^{-1/2}(C_5 + C_6 \Re^{-1}  + C_7\Rh^{-1})^{1/2}.
	\label{eq:bounds11}
\end{align}

%
\subsection{Negative $\alpha$}
We follow a similar approach for \vdd{negative} $\alpha$. In the statistically stationary state we have the generalised energy balance, $\epsilon_+ + \epsilon_- = \epsilon$ \vdd{discussed in Eq. \eqref{eq:energybal}, which can be similarly} derived from Eq. \eqref{eq:bounds1}, 
where \revi{$\epsilon_+ = -\nu_+ \avg{\psi \nabla^2q}$}, $\epsilon_- = \nu_-\avg{\psi q}$ and $\epsilon = \avg{\psi f_q}$. Using the generalised energy balance, we can write
\begin{equation}
  \epsilon_+ = \avg{\psi f_q} - \nu_-\avg{\psi q}
  \label{eq:enerbal}
\end{equation}
Using the Cauchy-Schwarz inequality to the first term  we get,
\begin{align}
	\epsilon_+ &\leq \avg{|\psi|^2}^{1/2} \avg{|f_q|^2}^{1/2} -\nu_-\avg{\psi q}\nonumber \\
 &\leq \avg{|\psi|^2}^{1/2} \avg{|f_q|^2}^{1/2}
	\label{eq:bounds12}
\end{align} 
where $F_q = \avg{|f_q|^2}^{1/2}$ is another measure of the forcing amplitude related to $F_\psi$ and \vdd{the large scale dissipative term $\nu_- \avg{\psi q}$} is positive definite, so it can be removed from the upper bound estimate in Eq. \eqref{eq:bounds12}. We can \vdd{then} apply Poincar\'e inequality to the first term as $\psi$ is a lower derivative than $\bold u$ and write
\revi{
\begin{align}
  \avg{|\psi|^2}^{1/2} &\leq C k_f^{-1}\avg{ |\bold u|^2}^{1/2} \nonumber \\
  &\leq C k_f^{-1} U,
  \label{eq:psi_rms2}
\end{align} }
\revi{where $C$ is a constant which depends on the domain geometry and the forcing function}. Combining Eqs. \eqref{eq:bounds12} and \eqref{eq:psi_rms2}, we can write 
\begin{align}
   \epsilon_+ &\leq C k_f^{-1} U F_q.
   \label{eq:ener_q}
\end{align}
Now, we multiply Eq. \eqref{eq:bounds1} by a smoothly varying, time independent function $\phi$, to give 
\begin{equation}
	\phi \partial_t q + \revi{\phi J(q, \psi)} = \nu_+ \phi \nabla^2 q -\nu_-\phi q + \phi f_q.
	\label{bounds13}
\end{equation}
By performing spatiotemporal averaging and changing the order of differentiation \vdd{using} integration by parts, we get 
\begin{equation}
	\vdd{\avg{\phi J(q, \psi)} - \nu_+ \avg{q \nabla^2 \phi} + \nu_-\avg{\phi q} = \avg{\phi f_q}}
	\label{eq:bounds14}
\end{equation}
as $\avg{\phi\partial_t q} = 0$ \vdd{due to statistical stationarity. 
We can use the Fourier space relation \eqref{eq:coupling} between $q$ and $\psi$ along with \eqref{eq:psi_rms} to write}
\begin{align}
\avg{|q|^2}^{1/2} &\leq \tilde{C}k_f^{\alpha}\avg{|\psi|^2}^{1/2}\nonumber \\
&\leq \tilde{C}k_f^{\alpha-1}U,
\label{eq:bounds14a}
\end{align}
\revi{where $\tilde{C}$ is another constant that depends on the domain geometry and the forcing function}. We now look to bound each term on the \vdd{left hand side} of Eq. \eqref{eq:bounds14} separately. For the second and third term, 
we can use Cauchy-Schwarz inequality and Eq. \eqref{eq:bounds14a} to get 
\begin{align}
  -\nu_+ \langle q\nabla^2\phi\rangle &\leq \nu_+ \avg{|q|^2}^{1/2} \avg{|\nabla^2\phi|^2}^{1/2} \nonumber \\
	&\leq \nu_+ \tilde{C} k_f^{\alpha-1} U \avg{|\nabla^2\phi|^2}^{1/2},
	\label{eq:bounds16}  
\end{align}
and
\begin{align}
  \nu_-\avg{\phi q}&\leq \nu_-\avg{|\phi|^2}^{1/2}\avg{|q|^2}^{1/2}\nonumber \\
  &\leq \nu_-\avg{|\phi|^2}^{1/2} \tilde{C} k_f^{\alpha-1}U.
  \label{eq:bounds17}
\end{align}
For the first term on the left hand side of \eqref{eq:bounds14}, we use the H\"older's and Cauchy-Schwarz inequalities \revi{along with} Eq.\eqref{eq:bounds14a} to get
\begin{align}
	\revi{\avg{\phi J(q, \psi)}=-\avg{(\bold u \cdot \nabla \phi) q}} &\leq ||\nabla\phi||_\infty \avg{|q|^2}^{1/2} \avg{|\bold u|^2}^{1/2} \nonumber \\
	&\leq ||\nabla\phi||_\infty \tilde{C} k_f^{\alpha-1}U^2,
	\label{eq:bounds18}
\end{align}

Putting equations \eqref{eq:bounds16},\eqref{eq:bounds17} and \vbj{\eqref{eq:bounds18}}  back in \eqref{eq:bounds14}, we get
\begin{equation}
	\avg{\phi f_{q}} \leq \tilde{C} ||\nabla\phi||_\infty k_f^{\alpha-1}U^2 +  
	\tilde{C} \nu_+ \vbj{k_f^{\alpha-1}} U\avg{|\nabla^2\phi|^2}^{1/2} + 
	\tilde{C} \nu_-\avg{|\phi|^2}^{1/2} k_f^{\alpha-1} U.
	\label{bounds17}
\end{equation}
Now, by letting $\phi = f_q/F_q$ we can simplify the expression in a similar manner \vdd{to} positive $\alpha$ and obtain a bound relating the forcing amplitude $F_q$ with the \vdd{rms velocity, viz.}
\revi{
\begin{equation}
	F_q \leq k_f^{\alpha} (\vbj{\tilde{C}_{1}^{'}} U^2 + \vbj{\tilde{C}_{2}^{'}} \nu_+ k_f U \vbj{+ \tilde{C}_{3}^{'}\nu_-k_f^{-1}U}),
	\label{bounds18}
\end{equation}
}
where \vbj{$\tilde{C_1}^{'}$, $\tilde{C_2}^{'}$  and $\tilde{C_3}^{'}$} are constants that depend on the form of the forcing function and the domain geometry. 
Substituting \eqref{bounds18} in the expression for the generalised energy dissipation rate \eqref{eq:ener_q} we obtain the following bound
\revi{
\begin{align}
	\epsilon_+ &\leq Ck_f^{\alpha -1} U (\vbj{\tilde{C}_1^{'}}U^2 + \tilde{C}_2^{'} \nu_+ k_f U + \tilde{C}_3^{'}\nu_-k_f^{-1}U),
	\label{bounds19}
\end{align}
}
which in non-dimensional form can be written as,
\begin{align}
	c_\epsilon &\leq \vbj{\tilde{C}_1} + \tilde{C}_2\Re^{-1} + \tilde{C}_3\Rh^{-1},
	\label{bounds20}
\end{align}
obtained after dividing by $k_f^{\alpha - 1} U^3$ and using the definitions \eqref{eq:c_eps_c_xi_m} and \eqref{eq:Reynolds_m}.

We then look to bound the generalised enstrophy dissipation rate $\xi_+= \nu_+ \avg{|\nabla q|^2}$, which can be written in the spectral space as
\begin{align}
	\xi_+ &= \nu_+ \sum_{\bf k} k^{1+\frac{\alpha}{2}} k^{1-\frac{\alpha}{2}} |\widehat q (\bold k, t)|^2.
\end{align}
Using the Cauchy-Schwarz inequality we get the expression
\begin{align}
	\xi_+ &\leq \nu_+ \left(\sum_{\bf k} k^{2+\alpha} |\widehat q (\bold k, t)|^2 \right)^{\frac{1}{2}} 
	\left(\sum_{\bf k} k^{2-\alpha} |\widehat q (\bold k, t)|^2 \right)^{\frac{1}{2}}.
	\label{eq:bounds21}		
\end{align}
For the first term in the brackets we can use \vdd{Eq. \eqref{eq:coupling} along with the inequality $k^{\alpha} \leq C' k_f^{\alpha}$, valid for negative values of $\alpha$, to get}, 
\begin{equation}
	\left(\sum_{\bf k} k^{2+\alpha} |\widehat q (\bold k, t)|^2 \right)^{\frac{1}{2}} = 
	\left(\sum_{\bf k} k^{2+3\alpha} |\widehat \psi (\bold k, t)|^2 \right)^{\frac{1}{2}}
	 \leq \vbj{\tilde{C}_{4} k_f^{3\alpha/2} U},
	\label{eq:bounds22}
\end{equation}
where $\tilde{C}_4$ is a constant that depends on the form of the forcing function and the domain geometry. 
Now, the second term in the brackets of \vdd{Eq. \eqref{eq:bounds21}} can be written as 
\begin{align}
	\left(\sum_{\bf k} k^{2-\alpha} |\widehat q (\bold k, t)|^2 \right)^{\frac{1}{2}} = 
	\avg{q\nabla^2 \psi}^{1/2} = (\epsilon_+/\nu_+)^{1/2},
	\label{eq:bounds23}
\end{align}
\vdd{using again Eq. \eqref{eq:coupling}.}
Putting equations \eqref{eq:bounds22} and \eqref{eq:bounds23} back in \eqref{eq:bounds21}, we get
\begin{align}
	\xi_+ &\leq \vbj{\tilde{C}_{4}} \nu_+^{1/2} k_f^{3\alpha/2} U \epsilon_+^{1/2}.
	\label{bounds24}
\end{align}
Using the bound for $\epsilon_+$ from Eq. \eqref{bounds19}, this inequality simplifies to 
\begin{align}
	c_\xi \leq \Re^{-1/2}(\vbj{\tilde{C}_{5}} + \vbj{\tilde{C}_{6}} \Re^{-1} \vbj{ + \tilde{C}_{7}\Rh^{-1}})^{1/2} 
	\label{bounds20}
\end{align}
in non-dimensional form.

\bibliographystyle{jfm}
\bibliography{citations}

\end{document}